\documentclass[num-refs]{wiley-article}

\usepackage{siunitx}

\usepackage{tikz}
\usepackage{tikz-qtree}
\usetikzlibrary{shapes}
\usetikzlibrary{trees}

\usepackage{booktabs}
\usepackage{lscape}
\usepackage{rotating}

\def\checkmark{\tikz\fill[scale=0.4](0,.35) -- (.25,0) -- (1,.7) -- (.25,.15) -- cycle;} 
\def\dash{\tikz\draw[thick][scale=0.4](0,0) -- (.7,0);}

\papertype{Original Article}
\paperfield{Journal Section}

\title{Container-based Cluster Orchestration Systems: A Taxonomy and Future Directions}

\author[1\authfn{1}]{Maria A. Rodriguez}
\author[1\authfn{1}]{Rajkumar Buyya}

\affil[1]{CLOUDS Laboratory, School of Computing and Information Systems, The University of Melbourne, Australia}

\corraddress{Maria A. Rodriguez}
\corremail{marodriguez@unimelb.edu.au}

\runningauthor{Rodriguez and Buyya}

\begin{document}

\maketitle

\begin{abstract} Containers, enabling lightweight environment and performance isolation, fast and flexible deployment, and fine-grained resource sharing, have gained popularity in better application management and deployment in addition to hardware virtualization. They are being widely used by organizations to deploy their increasingly diverse workloads derived from modern-day applications such as web services, big data, and IoT in either proprietary clusters or private and public cloud data centers. This has led to the emergence of container orchestration platforms, which are designed to manage the deployment of containerized applications in large-scale clusters. These systems are capable of running hundreds of thousands of jobs across thousands of machines. To do so efficiently, they must address several important challenges including scalability, fault-tolerance and availability, efficient resource utilization, and request throughput maximization among others. This paper studies these management systems and proposes a taxonomy that identifies different mechanisms that can be used to meet the aforementioned challenges. The proposed classification is then applied to various state-of-the-art systems leading to the identification of open research challenges and gaps in the literature intended as future directions for researchers.


\end{abstract}

\section{Introduction}

Containers have gained significant attention in recent years. They are standalone, self-contained units that package software and their dependencies together and provide process isolation at the operating system level.  Hence, similar to Virtual Machines (VMs), containers are a virtualization technique that enable the resources of a single compute node to be shared between multiple users and applications simultaneously.  However, instead of virtualizing resources at the hardware level as VMs do, containers do so at the operating system (OS) level. 

There are multiple technologies that realize the concept of containers. Perhaps the most widely used one is Docker~\cite{dockerOnline} but there are several products on the market including LXC~\cite{lxcOnline}, OpenVZ~\cite{openvzOnline}, Linux-VServer~\cite{vserverOnline}, and rkt~\cite{rktOnline}. Although with different underlying architectures and designed for different operating systems (e.g. Docker Windows containers vs. Docker Linux containers), there are various defining characteristics of containers that are a common denominator between different solutions. Firstly, containers executing on a single host share the operating system's kernel and run as isolated processes in user space, hence there is no need for a hypervisor. This isolation is done in such a way that there is no interference between applications, albeit some performance interference due to co-located processes competing for resources. This however is controlled to some extent by container managers by limiting the amount of resources such as CPU and memory that a container can use. Secondly, containers use as many system resources as they need at any given point in time and hence there is no need to permanently allocate resources such as memory. Finally, containers are spawned from images, which are executable packages that include everything that is needed to run them. This includes code, libraries, settings, and system tools. More importantly, these images can be constructed from filesystem layers and hence are lightweight and use considerably less space than VMs. 

As a result of the aforementioned features, containers provide a flexible environment in which applications are isolated from each other and offer benefits in terms of ease of deployment, testing, and composition to developers. Furthermore, they enable a better utilization of resources and the performance overhead that results from running applications in containers has been shown to be marginal by various studies~\cite{felter2015updated, morabito2015hypervisors, ruiz2015performance}. Their provisioning time has also been found to be much faster than VMs and in many cases, almost immediate~\cite{piraghaj2017containercloudsim}. These benefits have led to a considerable increase in the adoption and popularity of this technology. Containers are being widely used by organizations to deploy their increasingly diverse workloads derived from modern-day applications such as web services, big data and IoT in either private or public data centers. This in turn, has led to the emergence of container orchestration platforms. Designed to manage the deployment of containerized applications in large-scale clusters, these systems are capable of running hundreds of thousands of jobs across thousands of machines.  

Such orchestration systems are commonly designed to schedule a workload of containerized applications of one or more types. Each application type has its own characteristics and requirements such as high-availability long-running jobs, deadline-constrained batch jobs, or latency-sensitive jobs for instance. The majority of systems support multi-tenancy, that is, they schedule applications belonging to multiple users on a shared set of compute resources, allowing for better resource utilization. Hence, as applications are submitted for deployment, the orchestration system must place them as fast as possible on one of the available resources while considering its particular constraints and maximizing the utilization of the compute resources in order to reduce to the operational cost of the organization. These systems must also achieve this while handling a considerably large number of compute resources, providing fault tolerance and high-availability, and promoting a fair resource allocation. 

In summary, to achieve their goal, container orchestrating tools must efficiently manage a wide range of containerized applications and the distributed resources that support their execution. This is a challenging problem considering several issues that must be addressed such as scaling to a large number of machines, maximizing the application throughput, minimizing the application deployment delay, maximizing the resource utilization, meeting the specific requirements and constraints of different applications, providing fault tolerance and high availability, supporting different types of applications, and achieving a fair allocation of resources, among others. In this work, we aim to study how different container orchestration systems achieve these requirements as well as the different capabilities they offer. In the context of cloud computing, a similar problem has already been faced by VM resource managers, which are responsible for allocating compute, storage, and networking resources to applications within a data center. There are various surveys that explore this topic in great detail~\cite{jennings2015resource, manvi2014resource, mann2015allocation} and aid in understanding the foundations of the systems studied in this work, which can be seen as the evolution of VM-based infrastructure-as-a-service resource managers.

The rest of this paper is organized as follows. Section \ref{referenceArchitecture} presents a reference architecture for container-based cluster management systems. Section \ref{taxonomy} introduces the proposed taxonomy from three different perspectives, the application, scheduling, and resource models of cluster management systems. Section \ref{survey} describes and classifies various state-of-the-art systems followed by future directions in Section \ref{futureDirections} and a summary to conclude in Section \ref{summary}.

\section{Reference Architecture for Container Orchestration Systems}
\label{referenceArchitecture}

\begin{figure}[t!]
\centering
\includegraphics[scale=0.6]{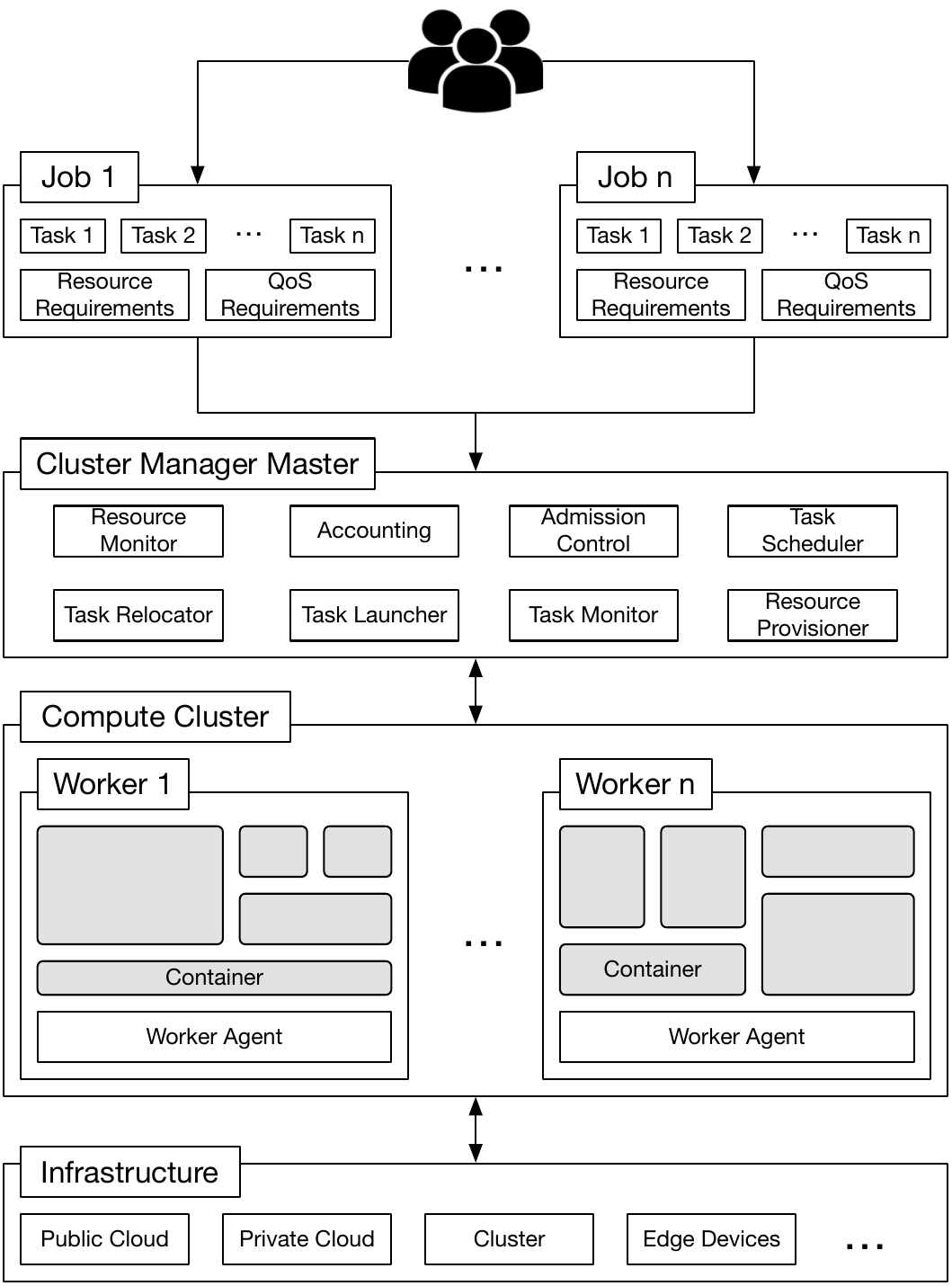}
\caption{A high-level container orchestration system reference architecture.}
\label{refArch}
\end{figure}

Container Orchestration Systems enable the deployment of containerized applications on a shared cluster. They enable their execution and monitoring by transparently managing tasks and data deployed on a set of distributed resources. A reference architecture is shown in Figure \ref{refArch}; the components shown are common to most container orchestration systems, however, not all of them have to be implemented in order to have a fully functional system. Four main entities or layers are identified in the presented architecture, namely one or more \emph{Jobs}, a \emph{Cluster Manager Master}, a \emph{Compute Cluster}, and the physical \emph{Infrastructure}. From a high-level perspective, users submit jobs composed of one or more tasks to the cluster manager master. This entity then assigns the submitted tasks to worker nodes in the compute cluster, where they are executed. The compute cluster is an abstraction of interconnected nodes that can be either physical or virtual machines on different infrastructures such as clouds or private clusters. A detailed explanation of each layer and its responsibilities is presented below.

\paragraph{Jobs} Users submit their applications in the form of jobs. These jobs usually belong to different users and are heterogeneous; they can range from long lasting latency sensitive services to short lived resource intensive batch jobs. A job is composed of one or more smaller \emph{tasks}. Tasks are generally homogeneous and independent, but some frameworks extend this definition and allow users to define jobs in terms of interdependent and heterogeneous tasks. Users can also express the \emph{resource requirements} of jobs in terms of the amount of CPU and memory they will require for example. Other \emph{QoS requirements} such as fault-tolerance requirements, time constraints, priorities, and QoS classes can be included as part of the job definition. 

\paragraph{Cluster Manager Master} The master component is the core of the orchestration system. It has a \emph{resource monitor} module responsible for keeping track of real-time resource consumption metrics for each worker node in the cluster. This information is usually accessed by other modules in the system as required. For instance, the task scheduler and relocator modules may use these data to make better optimization decisions. The \emph{accounting} module has a similar functionality to the resource monitor but focuses on collecting the actual resource usage and metrics relevant to the owner of the cluster management system. On one hand, infrastructure-related metrics consolidated by this module include the overall resource utilization, energy usage, and cost if deployed on a cloud environment. On the other hand, user-related metrics may include the number and type of jobs submitted by users as well as the amount of resources consumed by these jobs. These measurements may aid in enforcing user quotas or estimating billing amounts for example. 

The \emph{admission control} module is responsible for determining whether i) the user's resource quota is equal to or larger than the amount of resources requested or ii) there are enough resources available in the cluster to execute the submitted jobs. For the latter scenario, multiple decisions can be made in case resources are insufficient. On one side of the spectrum, jobs could simply be rejected. On the other side, a more complex solution would take into consideration users' priorities and QoS classes to preempt jobs deemed less important and free resources for those incoming jobs that are considered more important. Another possible solution would be to consider increasing the number of cluster nodes to place the incoming jobs.

The \emph{task scheduler} maps jobs, or more specifically tasks, onto the cluster resources. This is usually done by considering several factors and opposing goals. Firstly, the resource requirements and availabilities must be considered. Secondly, cluster management systems are concerned with efficiently using resources and hence, maximizing the utilization of the cluster nodes is usually an objective of schedulers. Finally, mapping tasks so that additional QoS requirements of jobs in terms of affinities, priorities, or constraints are met is another key scheduler responsibility.

The \emph{task relocator} can be seen as a rescheduler. Whenever tasks need to be relocated either because they are preempted or for consolidation purposes for example, this component is responsible for determining their fate. A task relocation policy may simply choose to discard the task or to place it back in the scheduling queue. More sophisticated approaches may analyze the state of the system and determine a new optimal location for the task with the aim of improving resource utilization. 

The \emph{task launcher} is responsible for launching the tasks' containers on specific cluster machines once this decision has been finalized by the scheduler or the relocator. Furthermore, to support the management of executing tasks, the \emph{task monitor} is responsible for auditing running tasks and monitoring their resource consumption and QoS metrics. This information aids in detecting failures or QoS violations and enables the system to make better scheduling or relocation decisions.

Finally, the \emph{resource provisioner} is in charge of managing the addition of new cluster nodes. This can be either a manual or an automatic process. In a manual process, usually system administrators will launch a new node with the worker agent software installed in it and execute a call for the agent to advertise itself to the master. This call can be processed by the resource provisioner so that the new node is now accounted for by the master. However, a resource provisioner is not always necessary in such a case, as the worker may automatically send a heartbeat signal to the resource monitor to advertise itself for instance. In the case of an automatic process however, the resource provisioner is an essential component of the architecture as it will be responsible for dynamically adding virtual nodes (i.e., virtual machines) to the cluster when the existing resources are insufficient to meet the applications' demands. It will also decide when nodes are no longer required in the cluster and will shut the nodes down to prevent incurring in additional costs. 

\paragraph{Compute Cluster} Each machine in the cluster that is available for deploying tasks is a worker node. Each of these nodes has a \emph{worker agent} with various responsibilities. Firstly, it collects local information such as resource consumption metrics that can be periodically reported to the master, specifically to the resource monitor. Secondly, it starts and stops tasks and manages local resources, usually via a container manager tool such as Docker or Linux Containers. Finally, it monitors the containerized tasks deployed on the node, information which is usually relied onto the task monitor component in the master.

\paragraph{Infrastructure} One of the main benefits of containers is their flexibility in being deployed in a multitude of platforms. Because of these, the cluster machines can be either virtual machines on public or private cloud infrastructures, physical machines on a cluster, or even mobile or edge devices among others. 

\section{Taxonomy}
\label{taxonomy}

In this taxonomy, we identify various characteristics and elements of container-based cluster orchestration systems. In particular, we study these platforms from the scheduling, application, and resource model perspectives. The aim of this section is to explain each taxonomy classification, examples and references to systems for each category are presented in Section \ref{survey}.

\subsection{Application Model}

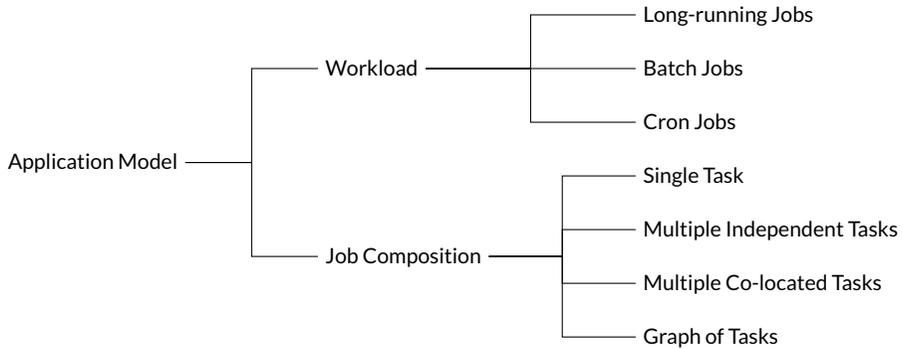
\begin{figure}
  \centering
\begin{tikzpicture}[grow=right]

\tikzset{level distance=120pt,sibling distance=2pt}
\tikzset{execute at begin node=\strut}
\tikzset{every tree node/.style={anchor=base west}}
\tikzset{edge from parent/.style= 
            {draw, edge from parent fork right}}
\Tree 
[.{Application Model} 
[.{Job Composition}     
    [.{Graph of Tasks} 
    ]
    [.{Multiple Co-located Tasks} 
    ]
     [.{Multiple Independent Tasks} 
    ]
     [.{Single Task}
    ]
]
[.{Workload}  
     [.{Cron Jobs} 
    ]
    [.{Batch Jobs} 
    ]
    [.{Long-running Jobs} 
    ]
]
]
\end{tikzpicture}
\caption{Application model taxonomy}
\label{appModelTaxonomy}
\end{figure}

In this section, we identify a classification for the application model used by container-based cluster management systems as depicted in Figure \ref{appModelTaxonomy}. This taxonomy is related to the \emph{Job} components as depicted in the reference architecture (Figure \ref{refArch}). In particular, we present various ways in which jobs or applications can be described by users as well as different characteristics that they can have.

To efficiently utilize resources, instead of running separate clusters of homogeneous containers, organizations prefer to run different types of containerized applications on a shared cluster. A common type of applications are long-running services that require high availability and must handle latency-sensitive requests. Examples include user-facing or web services. Another type of application are batch jobs. These have a limited lifetime and are more tolerable towards performance fluctuations. Examples include scientific computations or map-reduce jobs. Cron jobs are a type of batch jobs that occur periodically. The advantage for schedulers is that the time when cron jobs must be deployed is known in advance and hence this information can be used to make better scheduling decisions. 

Supporting a mixed workload may pose further challenges for container orchestrating systems as each type of application has different QoS requirements that must be fulfilled when performing the scheduling. However, having multiple applications share a cluster has significant benefits in terms of resource utilization. For example, Google \cite{verma2015large} demonstrated they would need 20-30\% more machines to run their workload if long running and batch jobs were segregated into separate clusters. Having a mixed workload allows for less performance-sensitive jobs to use resources that are claimed but not used by those with more stringent requirements for example. A batch job with a loose deadline could for instance tolerate being placed in a node with less CPU resources available than those requested for the job. As a result, better job packing can be achieved. 

Container orchestration systems may allow jobs to be defined by users in different ways. For instance, a job may be defined as a single containerized task. To offer more flexibility, frameworks generally allow jobs to be defined as a composition of multiple independent tasks that are identical, or almost identical to each other. Each of these tasks can then be deployed in a container on any given resource; this application model is used by Google's Borg~\cite{verma2015large} system for example. Another approach is to define a job in terms of multiple dependent tasks that must be co-located on the same node, as is done in Kubernetes~\cite{kubernetesOnline}.  Finally, systems like Apollo~\cite{boutin2014apollo} from Microsoft provide users with a more powerful application model in which they can define jobs as a graph of tasks. This enables users to define the communication patterns and dependencies between tasks and enables the scheduler to make better optimization decisions as more details on the application are known in advance.

\subsection{Scheduling}
\label{schedulingSubsection}

This section discusses containerized cluster management systems from the scheduling perspective. The classifications presented here relate mainly to the Task Scheduler and Task Relocator components from the \emph{Cluster Manager Master} entity described in Figure \ref{referenceArchitecture}.  In particular, we analyze the problem from two different angles, the first one is the scheduler architecture presented in Section 3.2.1 and the second one is related to the job scheduling policies depicted in Section 3.2.2.

\subsubsection{Scheduler Architecture}

\begin{figure}
  \centering
\begin{tikzpicture}[grow=right]
\tikzset{level distance=120pt,sibling distance=2pt}
\tikzset{execute at begin node=\strut}
\tikzset{every tree node/.style={anchor=base west}}
\tikzset{edge from parent/.style= 
            {draw, edge from parent fork right}}
\Tree 
[.{Scheduler Architecture}   [.{Two-level} 
     	]
	[.{Request-based} 
     	]
    ]
    [.{Decentralized} 
      	  [.{Monolithic} 
     	]
	[.{Modular} 
     	]
    ]
    [.{Centralized} 
  	  [.{Monolithic} 
     	]
     ]
]
\end{tikzpicture}
\caption{Scheduler architecture taxonomy}
\label{schedArchTaxonomy}
\end{figure}
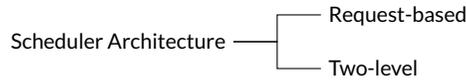

Regarding the scheduler architecture, there have been extensive studies on their classification on distributed systems. For example, Casavant and Kuhl~\cite{casavant1988taxonomy} proposed a taxonomy of scheduling algorithms in general purpose distributed computing systems, Toptal and Sabuncuoglu~\cite{toptal2010distributed} presented a classification of different factors related to distributed scheduling as well as a survey of the literature in the same topic, and Krauter et al.~\cite{krauter2002taxonomy} proposed a taxonomy and performed a survey of grid resource management systems for distributed computing that included the scheduler architecture or organization. Since this specific characteristic is key in understanding different approaches to scheduling in cluster management systems, we identify here the most prevalent architectures used by the surveyed frameworks as shown in Figure \ref{schedArchTaxonomy}. After briefly introducing their general definition, we aim to keep the discussion as relevant to the problem addressed in this work as possible. 

In a centralized architecture, there is a single scheduler responsible for making placement decisions for the containerized applications. These schedulers are monolithic in that they implement all the policy choices for the different types of workloads in a single code base. They have a global view of the system and the available resources and hence have the capability of choosing any of the existing nodes when making a placement decision. This also enables such schedulers to make better optimization decisions. However, they have the disadvantage of being a single point of failure in the system and suffering from scalability issues as the load of incoming scheduling requests increases and the number of nodes in the cluster grows. 

Decentralized architectures can be used to improve scalability. In this case, multiple distributed scheduler replicas exist. The replica instances can be monolithic, meaning that they handle a subset of the requests but implement all the policies and handle all the workloads. On the other hand, the replicas can be modular and hence, each instance can specialize on a specific application type or implement a different set of policies. For example, there may be  a scheduler for long running jobs and a scheduler for batch jobs. Each of these can in turn be replicated.  There are two key aspects that need to be considered by decentralized schedulers. The first one is determining how the requests are partitioned between schedulers. The second one consists on managing the state of the system between different replicas. For monolithic decentralized schedulers for example, the load can be partitioned using a traditional load balancing mechanism; for modular ones, the type of application to schedule will achieve this goal. 

Regarding state management, an approach is to provide each replica with access to the entire cluster state. If paired with optimistic concurrency control, despite the need to redo some work, this method successfully increments the parallelism of the schedulers. Omega~\cite{schwarzkopf2013omega} is an example of a system implementing a decentralized modular scheduler with shared state and optimistic concurrency control. On the other hand, each scheduler replica may have a partial view of the system state in which a particular resource is only made available to a particular scheduler at a time. Also known as pessimistic concurrency, this approach ensures there are no conflicts between schedulers by selecting the same resource for different applications. Another option is to implement an optimistic concurrency approach, in which all the schedulers have access to a shared state, increasing parallelism but also the potential for wasted scheduling effort if conflicts happen too often. 

Finally, in a two-level architecture, the resource management and the application framework are decoupled, and the scheduling is done in two separate layers. The bottom layer is responsible for managing the cluster resources and either offering available resources (i.e., offer-based) or granting resource requests (i.e., request based) to application frameworks. These application frameworks are then responsible for making the actual placement decisions, that is, determining which tasks will be deployed on which resources. This approach offers a great deal of flexibility to frameworks and mitigates the load and stress on the central scheduler. Mesos~\cite{hindman2011mesos} is an example of an offer-based two-level scheduler while Fuxi~\cite{zhang2014fuxi} is an example of a request-based one.  

It is worth mentioning that offer-based schedulers that hold a lock on resources offered to an application framework (i.e., pessimistic concurrency control) are more suited to application frameworks capable of making fast scheduling decisions on small and short-lived and small tasks ~\cite{schwarzkopf2013omega}. Mesos for example, alternates offering all the available cluster resources to each application scheduler.  As a result, long scheduling decisions by application frameworks would result in nearly all resources being locked and out of access to other schedulers during this period of time.

\subsubsection{Job Scheduling}

\begin{figure}
  \centering
\begin{tikzpicture}[grow=right]
\tikzset{level distance=120pt,sibling distance=2pt}
\tikzset{execute at begin node=\strut}
\tikzset{every tree node/.style={anchor=base west}}
\tikzset{edge from parent/.style= 
            {draw, edge from parent fork right}}
\Tree 
[.{Job Scheduling}   
    [.{Placement Constraints}          [.{Label-based} 
    ]
     [.{Limit-based} 
     ]
     [.{Query-based} 
    ]
      [.{Value-based} 
    ]
        [.{Affinity-based} 
    ]
    ]
    [.{Task Rescheduling} 
    ]
        [.{Task Preemption} 
    ]
       [.{Node Selection} 
	[.{Cluster Partition} 
     	]
	[.{Randomized Sample} 
     	]
	[.{All Nodes} 
     	]
     ]   
     ]
]
\end{tikzpicture}
\caption{Job Scheduling taxonomy}
\label{jobSchedTaxonomy}
\end{figure}
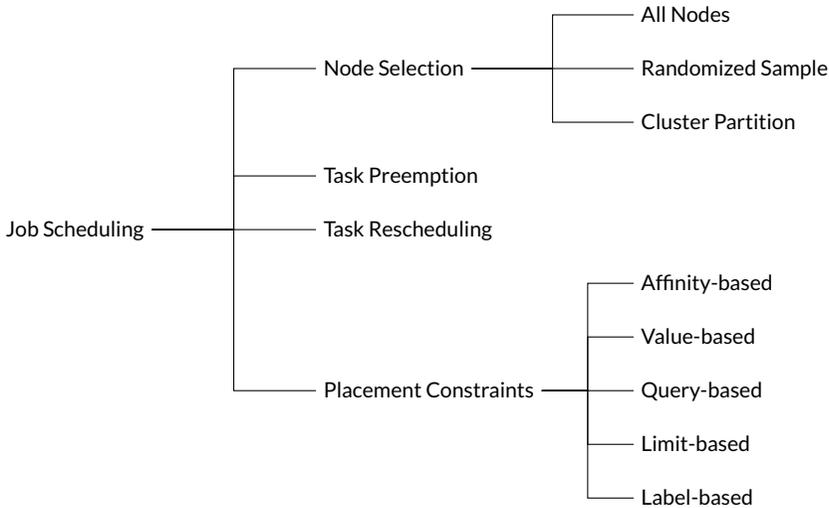

From the perspective of job scheduling, there are various policies and considerations that can be implemented by cluster managers. Namely, we consider in this work the constraints used to determine the placement of a task in a node, the pool of nodes that is considered when making this placement decision, whether tasks are preempted or not and whether rescheduling tasks is a feature supported by the system or not. This taxonomy is shown in Figure \ref{jobSchedTaxonomy}.

To improve the scalability of the system and to reduce the time from job submission to job placement, some schedulers will select resources to execute tasks from a subset of the cluster nodes, as opposed to evaluating the suitability of every single cluster node. This will speed up the decision time and enable the system to process more requests per time unit and to reduce the amount of time jobs have to wait before being assigned to resources. A possible strategy to achieve this is to select the best suited node to place a particular containerized task based on a randomized sample of the entire cluster nodes as done by Borg. Another approach is to partition the cluster into smaller sub-clusters and assign them to different scheduler replicas.

Schedulers can have greater flexibility in placing tasks if they can preempt existing assignments, as opposed to simply considering resources that are idle, but this comes at the cost of wasting some work in the preempted tasks. Preemption is useful when a priority scheme is in place. In this way, high priority tasks can obtain resources at the expense of lower priority ones, which are preempted. The system may even notify tasks before preempting them so that they have time to save their state and gracefully finish their execution. Preempted tasks will often be rescheduled elsewhere in the cluster though. This is one scenario in which rescheduling is currently being used in existing systems such as Borg. 

Another rescheduling use case is related to failures; when a worker node is deemed to have failed or is unreachable, the orchestrator will reschedule the tasks that were running on the machine on other nodes. Other systems like Kubernetes~\cite{kubernetesOnline} transparently manage the replication of failure-sensitive tasks via a replication controller. In this way, when a task terminates abruptly causing the number of current replicas to be smaller than the expected number of replicas, the task will be relaunched or rescheduled by the system. Finally, it is not unusual for tasks to be evicted from a node and rescheduled if they have exceeded their expected resource usage. This is further discussed in Section \ref{resManagement}.

Placement constraints are offered as a mean for users to customize the behavior of the scheduler to meet their applications' specific requirements. Hence, this feature does not apply to two-level schedulers, which leave the placement decisions to the application frameworks. For centralized and decentralized schedulers, value-based constraints are the  simplest way of achieving this with a user specified value matching a specific attribute of the node where the task is to be placed. Specifying the name of the node where a task must be deployed is an example of this type of constraint. Another example is specifying a hardware component that must be present in the host, such as the disk type being SSD. 

Query-based constraints on the other hand, offer more expression power to the users by enabling them to define more complex placement rules such as spreading the load evenly across a set of nodes and filtering attributes according to a set of values. For example, in Marathon~\cite{marathonOnline}, the constraint \emph{["rack\_id", "GROUP\_BY"]} will lead to tasks being evenly distributed across racks. 

Limit-based constraints allow the control of machine diversity, that is, they enforce per-attribute limits such as the number of tasks of a given type to be deployed on a single host. Their functionality is actually a subset of the functionality offered by query-based constraints. An example would be a constraint ensuring that no more than two instances of a job are deployed on a single host. Aurora~\cite{apacheAuroraOnline}, a framework built on top of Mesos implements this type of scheduling constraints. 

Label-based constraints are similar to value-based but offer more flexibility in that users can define their own labels used to specify identifying attributes of objects that are meaningful and relevant to them, but that do not reflect the characteristics or semantics of the system directly. For example, in Kubernetes labels can be used to force the scheduler to co-locate tasks from two different jobs that communicate a lot into the same availability zone.

Finally, affinity-based placement constraints enable users to define rules on how jobs can be scheduled relative to other jobs. An affinity rule would lead to jobs being co-located whereas an anti-affinity one would prevent jobs from being co-located. A key difference with the other placement constraints is that affinity rules apply constraints against other jobs running on a node, as opposed to applying constraints against labels or features of the actual node. Finally, it is worthwhile mentioning that the use of constraints can significantly impact the performance of scheduling algorithms. An affinity-based constraint in Yarn~\cite{vavilapalli2013apache} would enable users for example to place two containers with tag \emph{x} on a node on which containers with tag \emph{y} are running.

\subsection{Cluster Infrastructure and Management}
\label{resManagement}

This section describes different characteristics of the compute cluster, its management, and the underlying infrastructure supporting it. Section 3.3.1 presents a taxonomy for the cluster infrastructure, Section 3.3.2 introduces a taxonomy for different resource management techniques, Section 3.3.3 outlines different cluster-wide objectives, and finally Section 3.3.4 introduces different features of container-based cluster management systems that support multi-tenancy. 

\subsubsection{Cluster Infrastructure}
\label{clusterInfSection}

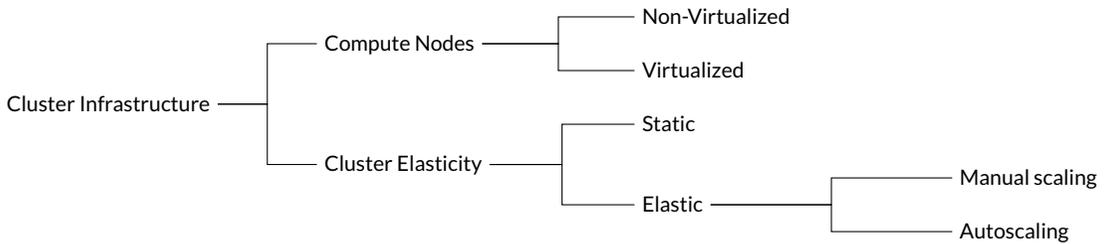
\begin{figure}
  \centering
\begin{tikzpicture}[grow=right]
\tikzset{level distance=120pt,sibling distance=2pt}
\tikzset{execute at begin node=\strut}
\tikzset{every tree node/.style={anchor=base west}}
\tikzset{edge from parent/.style= 
            {draw, edge from parent fork right}}
\Tree 
[.{Cluster Infrastructure}  
    	 [.{Cluster Elasticity}           [.{Elastic} 
          [.{Autoscaling} 
   	  ]
	  [.{Manual scaling} 
   	  ]
          ]
          [.{Static} 
         ]
         ]
          [.{Compute Nodes}  [.{Virtualized} 
   	 ]
	 [.{Non-Virtualized} 
          ]
	  ]
]
\end{tikzpicture}
\caption{Cluster Infrastructure taxonomy}
\label{infrastructureTaxonomy}
\end{figure}

Figure \ref{infrastructureTaxonomy} depicts the taxonomy for the cluster infrastructure. This classification defines the different types of infrastructures that can support the compute cluster, which based on their characteristics may have a significant impact on how the compute cluster resources are managed by the cluster manager master. 

The resources in a cluster may be static in that there is a fixed number of machines that remains relatively constant (failures, maintenance, and new hardware addition, may be exceptions) or elastic in that the cluster scales out and in based on the current demand of the system. Static settings are more likely to be based on bare metal, or non-virtualized servers. Their management consist on efficiently utilizing all of the existing resources. Elastic approaches on the other hand, expect the cluster to vary in size over time and are more common on virtualized environments. This scaling may be available through manual methods, in which a system administrator or an external framework adds (or removes) one or more compute resources to (from) the cluster, which then become available (unavailable) to the scheduler for placing containers. Autoscaling methods on the other hand, enable the addition and deletion of resources from the cluster in an automated fashion. The autoscaler is part of the system as opposed to an external decision maker and hence decides whether nodes should be added or removed based on the current state of the system. The current utilization of resources would be a good indicator to trigger such decision for example. Although the latter scenario considerably increases the complexity of the system, it may enable a better utilization of resources leading to decreased operational costs, a potential improvement in performance, and reduced energy consumption. \\

\subsubsection{Resource Management} 
\label{resManagementSection}

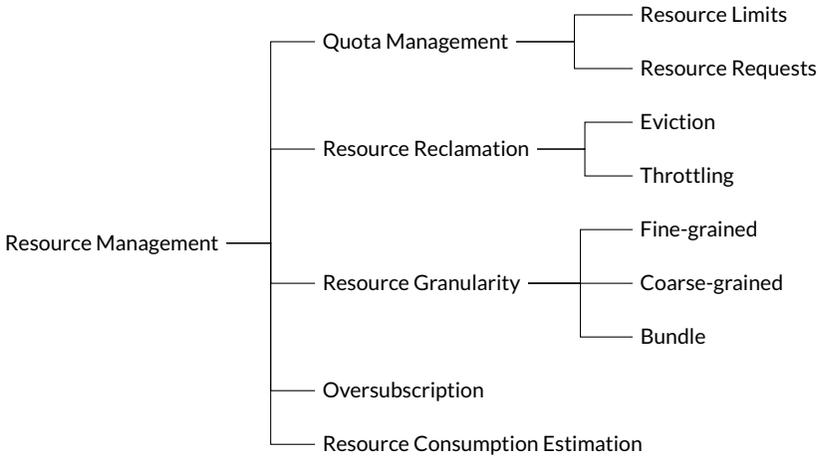
\begin{figure}
  \centering
\begin{tikzpicture}[grow=right]
\tikzset{level distance=120pt,sibling distance=2pt}
\tikzset{execute at begin node=\strut}
\tikzset{every tree node/.style={anchor=base west}}
\tikzset{edge from parent/.style= 
            {draw, edge from parent fork right}}
\Tree 
[.{Resource Management}  
	[.{Resource Consumption Estimation}
	]
   	[.{Oversubscription} 
    	]
	[.{Resource Granularity} 
		   [.{Bundle}
		   ]
	           [.{Coarse-grained} 
		   ]
		   [.{Fine-grained}
		   ]
	]
    	[.{Resource Reclamation}
		   [.{Throttling}
		   ]
	           [.{Eviction} 
		   ]
	]
         [.{Quota Management} 
  	 	[.{Resource Requests}
		]
		[.{Resource Limits}
		]
    	 ]
]
\end{tikzpicture}
\caption{Resource Management taxonomy}
\label{resMngmntTaxonomy}
\end{figure}

Different resource management techniques implemented by the \emph{Cluster Manager Master} are illustrated on Figure \ref{resMngmntTaxonomy}. Allocating computational resources in distributed environments is a challenging problem. Orchestration systems usually achieve this with task resource requests, which allow users to define the amount of resources such as memory and CPU that a given task will use (at its peak time). This information is then used by the framework to efficiently assign containers to machines based on their available resources and the tasks currently deployed on them. 

Resource limits on the other hand, are an upper bound on the amount of resources a job or a task is allowed to consume. They are usually enforced by orchestration systems by means of resource reclamation. That is, by either throttling the use of a given resource when a task has exceeded its limit or by evicting the task. The choice between throttling or killing generally depends on whether the over consumed resource is compressible or non-compressible. Compressible resources are those to which the amount used by a task can be controlled without the need of killing it, an example is CPU capacity. Non-compressible resources are those that cannot be reclaimed without killing the task, an example is RAM. It is worthwhile noticing that resource reclamation does not have to be enforced for every task in the system, contrary to this, many frameworks apply these methods selectively to tasks with specific priorities or importance within the system. For example, Borg will only throttle or kill non-production tasks, but would never apply these rules to production ones. 

Resource granularity refers to the way in which resources are allocated to tasks, which equates to the way in which tasks are allowed to express resource requests. Some systems assign tasks to fixed-sized coarse-grained slots. In this way, CPU cores are assigned for example in units of 1 and RAM in units of 256 bytes. An example of such system is YARN; which enables CPU requests to be made in increments of 1 virtual core and RAM requests in increments of 1 megabyte. These increment values can be configured by users, but they must be larger than the aforementioned ones. Fine-grained resource granularity on the contrary, gives users more control and flexibility on the amount of resources they request. Borg users for example request CPU in units of milli-cores and memory and disk space in bytes. According to the authors findings \cite{verma2015large}, this enabled the system to have a better resource utilization by requiring 30-50\% less resources in the median case than when rounding up resource requests to the nearest power of two. For the purpose of this taxonomy, we define units equal to or larger than one CPU core and 1 megabyte for RAM or disk to be coarse-grained, anything smaller we classify as fine grained. Yet another option is to bundle resources, just as virtual machines are. In this way, a resource unit is specified as a tuple of resource amounts and allocations are done in number of resource units. An example are \emph{ScheduleUnits} in Fuxi, defined as a unit size description of a set of resources such as <1 core CPU, 1GB Memory>. Resource requests and allocations are then specified in number of \emph{ScheduleUnits}.

To achieve a better utilization of the available resources, many frameworks oversubscribe their servers. This idea is based mainly on two observations. Firstly, it is uncommon for tasks to consume the amount of resources they requested throughout their entire lifetime and instead their average usage is usually significantly lower. Secondly, users tend to overestimate the amount of resources they request for a given task. Hence, rather than letting unused requested resources go unused, container orchestration systems may choose to assign these unused reserved resources to tasks that can tolerate lower-quality resources. To be able to achieve this safely, a relatively accurate estimate of the actual resource consumption of tasks must be made by the orchestration system. When tasks exceed this estimate, the resources must be reclaimed either by throttling or killing the opportunistic tasks which may eventually need to be rescheduled somewhere else. 

Resource consumption estimation is used to predict and estimate the amount of resources a container consumes at different points in time, as opposed to relying simply on the amount of resources requested for a particular container. The reason is twofold. Firstly, resource requests are usually misestimated, and overestimated, by users. Secondly, the resource consumption of a task is likely to vary over time, with the peak consumption spanning only over a fraction of its lifetime. Both scenarios lead to resources that are reserved but are idle most of the time and hence lead to the cluster being underutilized. By monitoring and estimating the resource consumption of containers then, better oversubscription and opportunistic scheduling decisions can be made by the system. 

\subsubsection{System Objectives}
\label{sysObjectivesSection}

\begin{figure}
  \centering
\begin{tikzpicture}[grow=right]
\tikzset{level distance=120pt,sibling distance=2pt}
\tikzset{execute at begin node=\strut}
\tikzset{every tree node/.style={anchor=base west}}
\tikzset{edge from parent/.style= 
            {draw, edge from parent fork right}}
\Tree 
[.{System Objectives}   	
		  [.{Application-specific QoS}
		  ]	
		  [.{High Scheduling Throughput}
		  ]
		   [.{High Availability}
		   ]
	           [.{High Resource Utilization} 
		   ]
		  [.{Scalability}
		  ]
	]
\end{tikzpicture}
\caption{System objectives taxonomy}
\label{systemObjectivesTaxonomy}
\end{figure}
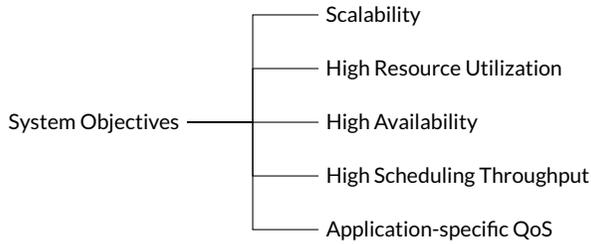

Many of the resource management mechanisms are put in place in order to fulfil a higher-level goal, referred to as the system objectives. These objectives ultimately guide the design and decisions made by the modules in the \emph{Cluster Management Master}. Their classification is outlined in Figure \ref{systemObjectivesTaxonomy}.

Scalability is a primary goal of existing systems. An approach to achieving it is to avoid a centralized scheduler. For example, having scheduler replicas and an optimistic concurrency management approach to sharing the cluster state seems to lead to high-levels of scalability. Making quick scheduling decisions will also have a positive impact on the scalability of the system. Other policies used for this purpose include caching the score given to nodes when selecting them in the scheduling process, using equivalence classes to group tasks with similar characteristics for scheduling purposes, and selecting a node to place a task from a subset of the cluster nodes~\cite{verma2015large}.

Ensuring that applications deployed on the cluster have high-availability is another goal for many systems. To achieve this, management systems must be able to appropriately handle, mitigate, and detect node and task failures. Some systems achieve this by automatically rescheduling tasks when they fail or are evicted, spreading task replicas across machines and racks, and avoiding redeploying tasks on a machine where the task previously failed, among others. 

Achieving high cluster utilization is also a primary concern, especially in proprietary clusters as under-utilizing resources will ultimately lead to higher operational costs for the organization. Oversubscription, fine-grained resource allocation, resource consumption estimation, resource reclamation, and preemption are some of the methods used for this purpose. 

Maximizing job throughput is another common goal of systems. This will eventually lead to a more scalable system and to a reduced delay between job submission and job placement. This may be crucial for applications requiring low response times or latencies or with a very strict deadline. Ultimately, high throughput is achieved by making fast scheduling decision and may  require the optimality of the placement decisions to be compromised. 

Finally, management systems may aim to satisfy Quality of Service (QoS) requirements that are specific to the applications they serve. This may be an inherent characteristic in two-level schedulers as the application frameworks can use the resources requested or offered to them in such a way that different objectives such as meeting a deadline or minimizing the makespan of a job are fulfilled. For systems with a centralized or decentralized built-in scheduler this goal is not as straightforward. Of the surveyed systems, Apollo is the only one that considers the applications' goals when scheduling. It aims to deploy tasks so that their makespan is minimized. It does however assume that this is a common QoS requirement across all the applications it handles; managing heterogeneous QoS parameters for different applications is an interesting challenge that has not been addressed yet. 

In this taxonomy, systems are classified as having one of the above objectives if and only if the functionality to achieve them is built-in into the core components of the system. For instance, the scheduler must have policies in place that aim to allocate tasks across nodes in different power domains to achieve high-availability. We do not consider the case in which a framework offers users the means of achieving these objectives by using mechanisms such as placement constraints, job replication, or QoS classes.  

\subsubsection{Multi-tenancy Features}
\label{multitenancySection}

\begin{figure}
  \centering
\begin{tikzpicture}[grow=right]

\tikzset{level distance=37mm,sibling distance=2pt}
\tikzset{execute at begin node=\strut}
\tikzset{every tree node/.style={anchor=base west}}
\tikzset{edge from parent/.style= 
            {draw, edge from parent fork right}}
\Tree 
[.{Multi-tenancy Features}  
 [.{Performance Isolation} 
     [.{Per-tenant Resource Quota} 
    ]
     [.{Compute Resource Isolation} 
    ]
    ]
     [.{Network Isolation} 
    [.{Third-party Network Isolation} 
    ]
     [.{IP-per-container} 
    ]
    [.{Port Mapping} 
    ]
    ]
    [.{Security Isolation}
    [.{Container-centric} 
     [.{Hypervisor Isolation} 
    ]
     [.{Restricted Execution} 
    ]
    ]
    [.{Access control} 
      [.{RBAC-based Authorization} 
    ]
      [.{Authentication} 
    ]
      [.{Restricted API Access} 
    ]
    ]
    ]
]
]
\end{tikzpicture}
\caption{Multi-tenancy features taxonomy}
\label{multitenancyTaxonomy}
\end{figure}
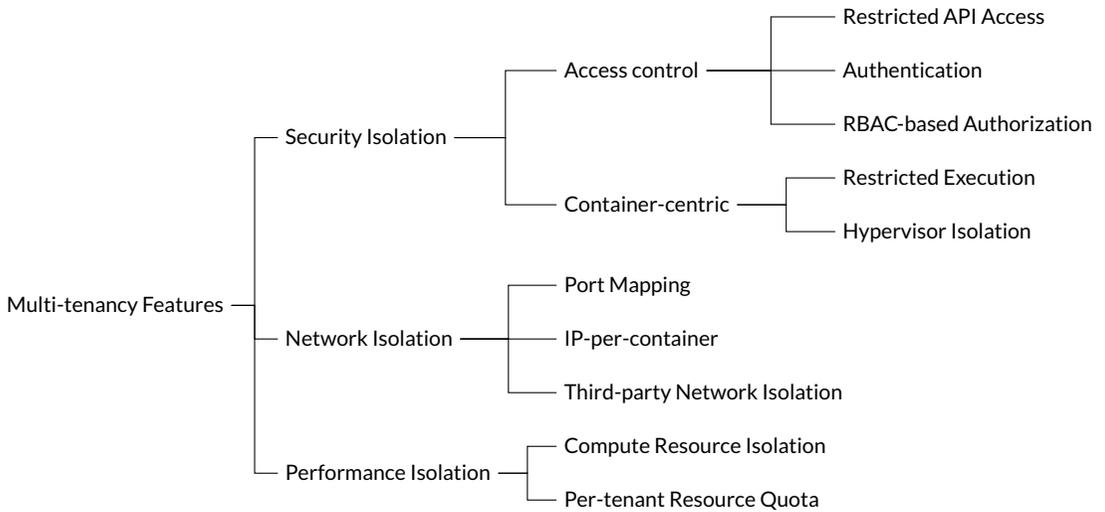

Multi-tenancy refers to the ability of a system or framework to serve multiple tenants in a physically shared environment. A tenant is defined as a user or an application that must be kept logically isolated from other tenants. Within the context of cluster orchestration systems, two particular scenarios arise based on the membership of tenants to either the same or different organizations. On one hand, multi-tenancy across different organizations enables a single cluster to be shared between users or stakeholders belonging to different companies. In this scenario, applications deployed on the cluster are untrusted and hence strong policies isolating tenants are necessary. For example, each tenant should have access to their own set of control plane objects (e.g., master component), resource quotas should be enforced based on how much tenants have paid, network isolation should be strong to prevent communication between applications belonging to different tenants, performance and security isolation of applications may require stronger guarantees than what is currently provided by containers (e.g. VMs), and sole-tenant nodes may be necessary in some particular instances. 

On the other hand, multi-tenancy within a single organization sees the resources of a cluster being dedicated to a particular company but shared between different employees or teams within that organization. Cluster management systems are commonly deployed under this scenario, which requires a less strict degree of isolation between tenants as it is assumed all applications deployed in the cluster are trusted. In this case, the cluster control plane can be shared among tenants, the current level of isolation provided by shared-OS containers is usually sufficient, and communication across tenancy domains may be desirable in some instances. 

Based on this, we have identified different multi-tenancy features as illustrated in Figure \ref{multitenancyTaxonomy} that are present in cluster management systems and provide different levels of isolation between tenants. Security isolation policies refer to features restricting access to the cluster resources and what containers can do on a particular node. In particular, we categorize frameworks based on whether they provide (or not) the following access control, isolation-enabling features. An authentication module capable of identifying genuine tenants, an API which allows tenants to create, access, or modify only those objects they own and hides those that belong to other tenants, and an authorization module that ensures tenants can only access those resources they are entitled to (e.g., role-based access control (RBAC) capabilities). Container-centric isolation defines whether a framework supports the definition of policies that restrict what the code executing inside containers can or cannot do (e.g., disable applications running as root) or whether hypervisor-based containers with hardware-enforced isolation are supported or not. Examples of hypervisor-based containers are Kata containers~\cite{kataOnline}, runV~\cite{runvOnline} and vSphere Integrated Containers~\cite{vsphereOnline}.

Network isolation features mask the use of a single physical network by applications from multiple tenants. They usually allow restrictions to be put in place defining the way in which applications across different tenant domains can communicate with each other. In an IP-per-container model, each container is assigned a unique IP address and as a result, owns the entire available port range. This model enable containers to resemble VMs or physical hosts from the networking perspective. Hence, rules can then be put in place defining address ranges to or from which applications are allowed to send or receive packets. This rule can be defined and enforced in different manners. For instance, a simple approach relying on network routing tables can be devised or third-party vendor plugins can be used to provide network isolation for different requirements. Examples of these third-party plugins include Calico~\cite{calicoOnline}, Flannel~\cite{flannelOnline}, and Nuage~\cite{nuageOnline}. 

Port mapping on the other hand, forces all of the containers running in a node to share the same IP. Containers on the node are assigned a separate network stack, usually via Linux network namespaces. In this way, each container is assigned a non-overlapping range of ports and only packets addressed to or from these ports are delivered. Other network stack resources that are partitioned include routes and firewall rules, which enables the network within a single running kernel instance to be virtualized.
 
Finally, performance isolation policies refer to features that prevent a tenant's resource usage from negatively impacting other tenant's applications. Achieving compute resource usage isolation is a key aspect, which is accomplished to some degree by all frameworks within the scope of this work by means of containerization. Other mechanisms that facilitate performance isolation include enabling and enforcing a per-tenant resource quota and scheduling related policies such as placement constraints and preemption. A discussion on these scheduling features was presented in Section \ref{schedulingSubsection}.

\section{Survey}
\label{survey}

This section discusses a set of state-of-the-art container orchestration systems and analyzes them in the context of the presented taxonomy. The results are summarized in Tables \ref{orgSurvey}, \ref{appModelSurvey}, \ref{rschedSurvey}, \ref{clusterInfSurvey}, \ref{resMngmntSurvey}, and \ref{sysObjSurvey}.

\subsection{Borg}
Google's Borg~\cite{verma2015large} cluster management system is designed to run hundreds of thousands of heterogeneous jobs across several clusters, each with tens of thousands of machines. Users submit jobs to Borg, which are composed of one or more homogeneous tasks. Each job runs in a cell, which is a set of heterogeneous machines managed as a unit. The workload in Borg cells is composed of two types of applications. The first are long running services that must remain available at all times. These services must serve short-lived requests with minimal latency and correspond mostly to end-user-facing web applications. They are commonly classed as high-priority or production jobs. The second type of workload corresponds to batch jobs. These can take from a few seconds to days to complete and are commonly classed as lower-priority or non-production jobs. Tasks run on containers deployed on physical machines and have resource requirements specified in terms of disk space, RAM , and CPU cores, among other resources. The container stack used is proprietary and is based on Linux cgroups~\cite{cgroupsOnline}. However, an open source version of this stack, called lmctfy~\cite{lmctfyOnline}, is readily available.

Regarding the scheduler, a queue of pending tasks is asynchronously monitored. This queue is transversed in a high to low priority order with jobs selected based on a round robin scheme within each priority. The scheduling algorithm has two parts: finding feasible machines that match the task's requirements and choosing one of these machines (scoring). The scoring mechanism favors machines that already have the tasks' packages, spreading tasks (from the same job) across power and failure domains, and packing quality like mixing low and high priority tasks on the same machine to allow the high priority ones to consume more resources when required (e.g., in a load spike).

Borg does not rely solely on the amount of resources requested for a task to reserve CPU and RAM for example. Instead, it estimates the amount of resources a task will use and reclaims the rest for work that can tolerate lower quality resources such as batch jobs. This reservation changes dynamically based on the fine-grained resource consumption of the task measured every few seconds. Borg differentiates between compressible resources like CPU cycles that can be reclaimed from a task by decreasing its QoS without killing it; and non-compressible resources like memory, which generally cannot be reclaimed without killing the task. Hence, tasks that try to consume more ram are killed, while CPU is throttled to the requested amount. 

Borg uses various technologies as building blocks to achieve its goals. For instance, Chubby~\cite{burrows2006chubby}, a distributed lock service implemented at Google that provides strong consistency, supports the framework's name system that allows tasks to be located by clients and other services. To enable the monitoring of tasks, these have a built-in HTTP server that publishes information about their health as well as different performance metrics. Supporting the accounting module by recording data such as job submissions and task resource usage is Infrastore, a scalable read-only data store with an interactive SQL-like interface via Dremel~\cite{melnik2010dremel} also developed at Google. A Paxos-based store~\cite{lamport1998part} is used to maintain the replicated state of five master processes in a highly available manner. Finally, Linux \emph{chroot} jail is used as part of the container stack to securely isolate multiple tasks running on the same machine.

Overall, Borg is one of the most advanced surveyed systems. It supports features such as oversubscription with resource consumption estimation, mixed workloads, fine-grained resource requests, and task preemption among others. Its cluster utilization rate was reported to be between 60 and 70 percent and its throughput to be approximately 10 thousand tasks per minute in a cluster composed of tens of thousands of nodes. 

\subsection{Kubernetes}
Kubernetes~\cite{kubernetesOnline} is a framework designed to manage containerized workloads on clusters. The basic building block in Kubernetes is a pod. A pod encapsulates one or more tightly coupled containers that are co-located and share the same set of resources. Pods also encapsulate storage resources, a network IP, and a set of options that govern how the pod's container(s) should run. A pod is designed to run a single instance of an application; in this way multiple pods can be used to scale an application horizontally for example. The amount of CPU, memory, and ephemeral storage a container needs can be specified when creating a pod. This information can then be used by the scheduler to make decisions on pod placement. These compute resources can be specified both as a requested amount or a as cap on the amount the container is allowed to consume. 

The scheduler ensures that the total amount of compute resource requests of all pods placed in a node does not exceed the capacity of the node. This even if the actual resource consumption is very low. The reason behind this is to protect applications against a resource shortage on a node when resource usage later increases (e.g., during a daily peak).  If a container exceeds its memory limit, it may be terminated and may be later restarted. If it exceeds its memory request, it may be terminated when the node runs out of memory. Regarding the CPU usage, containers may or may not be allowed to exceed their limits for periods of time, but they will not be killed for this. On the other hand, containers and pods that exceed their storage limit will be evicted. Other resources (called Extended Resources) can be specified to advertise new node-level resources, their resource accounting is managed by the scheduler to ensure that no more than the available amount is simultaneously allocated to pods.

From a technical perspective, Kubernetes allows for various types of container runtimes to be used, with Docker and rkt natively supported by the platform. More recently, the release of the framework's Container Runtime Interface (CRI) API has enabled Kubernetes to support other container technologies such as containerd~\cite{containerdOnline} and frakti~\cite{fraktiOnline}, a hypervisor-based container runtime. Furthermore,  CRI-O~\cite{crioOnline}, an implementation of the CRI API, currently enables Kubernetes to support any OCI (Open Container Initiative) compliant container runtime such as runc~\cite{runcOnline}. Also, supporting Kubernetes in managing the cluster nodes and jobs is etcd~\cite{etcdOnline}, an open source, highly-available, distributed key-value store. Specifically, etcd is used to store all of the cluster's data and acts as the single source of truth for all of the framework's components.

Overall, Kubernetes is a highly mature system; it stemmed from ten years of experience at Google with Borg and Omega and is the leading container-based cluster management system with an extensive community-driven support and development base. It provides users with a wide range of options for managing their pods and the way in which they are scheduled, even allowing for pluggable customized schedulers to be easily integrated into the system. It is worthwhile noticing that the Kubernetes built-in scheduler is classified as decentralized and monolithic in this survey, however, this can be overridden by the use of pluggable schedulers which, based on their implementation, can be either centralized, monolithic decentralized, or modular decentralized. To conclude, although Kubernetes' performance and scalability may still not reach the levels of industry-based systems like Borg, as of version 1.10, Kubernetes is capable of supporting clusters of up to 5000 hundred nodes~\cite{kubernetesEvalOnline}, which suits the needs of many organizations nowadays. 

\subsection{Swarm}
Docker Swarm~\cite{dockerSwarmOnline} orchestrates applications on a cluster of physical or virtual nodes running Docker. Users can execute regular Docker commands (i.e., the standard Docker API) which are then executed by a swarm manager. The swarm manager is responsible for controlling the deployment and the lifecycle of Docker-based containerized applications on the cluster nodes. These applications are represented as services and may be composed of one or more containers. Notice however that these services are intended to be long-running jobs that are continuously running. The replication of services is allowed and is transparently handled by the swarm manager when an application is deployed. However, the failure of nodes or services is not automatically handled by the framework and hence replicas are not redeployed if they fail.  

Three scheduling strategies are available to deploy containers. \emph{Spread} selects the node with the least number of containers deployed on it, \emph{binpack} selects the node with the minimum amount of CPU and RAM available, and \emph{random} chooses a random node from the cluster. Swarm also supports placement constraints to enable users to customize the behavior of the scheduler based on their requirements. These include value, label and affinity-based constraints. In this way, containers can be deployed on nodes with a specific operating system, stored container image, deployed service, or that belong to a specific cluster rack for example. 

Overall, Swarm specializes on the scheduling of Docker containers, it is easy to use for Docker users and is lightweight and flexible. Furthermore, the simplicity of the scheduling policies leads to a scalable system where placement decisions are made fast; its performance in this matter has been corroborated by evaluating it an environment running 30,000 containers~\cite{swarmEval}. However, the system lacks some key functionality offered by more robust management systems like Kubernetes. For instance, Swarm does not handle failures of nodes, whereas the majority of surveyed systems are capable of automatically maintaining a specified number of container replicas running throughout an application's lifetime. 

\subsection{Mesos}

Mesos~\cite{hindman2011mesos} is an open-source management system built around the idea of two-level scheduling. It delegates the scheduling of tasks to application frameworks, such as Hadoop and MPI, while it remains responsible for distributing the available cluster resources among all of its client frameworks. In particular, scheduling in Mesos is done in two phases. In the first stage, Mesos divides the resources of the cluster and periodically presents each application framework with a resource offer. These resource offers are based on policies that preserve priorities and fairness. Frameworks can accept or reject these offers. Once an application framework accepts an offer, it can schedule its tasks on the obtained resources using its own scheduler. Afterwards, Mesos actually launches the tasks for the framework on the corresponding hosts.

As mentioned in Section \ref{schedArchTaxonomy}, offer-based two-level schedulers may encounter difficulties when application frameworks have long scheduling cycles and place mostly long-running tasks that tend to hog resources. To alleviate these issues, Mesos resource offers are time-bounded and hence applications are incentivized to make fast scheduling decisions. The reservation of resources for short tasks is also possible in Mesos. In this way, a specific set of resources on each node can be associated with a maximum task duration, time after which tasks running on those resources are killed. It could be argued that the benefits of this are twofold. First, it creates an incentive for applications to deploy short tasks, and second, it alleviates the impact that long tasks may have on the system.

Mesos provides application frameworks with high-availability by maintaining multiple replicas of the master component. ZooKeeper~\cite{hunt2010zookeeper} is used to elect a leader within the replicated masters. Regarding the container runtimes supported, Mesos allows users to deploy applications packed in Docker containers or in its custom, Linux-based containers called Mesos containers. 

Altogether, Mesos provides a large degree of flexibility to its users and applications. Users can deploy different types of frameworks on top of Mesos to suit their requirements. Applications on the other hand can apply their specialized knowledge and schedule applications based on it. Furthermore, the unique scheduling model of Mesos and its use of fine-grained resource sharing-model, enables it to achieve high-utilization while remaining adaptable to workload changes and robust. The simplicity of the system also allows it to scale to 50,000 nodes.

In the following sections, we introduce Aurora and Marathon, two orchestration frameworks that rely on Mesos to manage the cluster resources. Even though they do not implement the entire functionality of a cluster management system, they are included in the survey as they provide users with different application models and different scheduling features that complement those offered by Mesos.

\subsubsection{Aurora}
Originally developed by Twitter, Aurora~\cite{apacheAuroraOnline} is a scheduler that runs on top of Mesos and enables long-running services, cron jobs, and ad-hoc jobs to be deployed in a cluster. Aurora specializes in ensuring that services are kept running continuously and as a result, when machine failures occur, jobs are intelligently rescheduled onto healthy machines. Furthermore, as opposed to Mesos, Aurora handles jobs which are composed of multiple, near identical tasks. Each task is in turn composed of one or more processes, which are managed by an executor process that runs on worker nodes and is responsible for launching and monitoring tasks. To deploy a job in Aurora, a job configuration is first submitted. This configuration specifies the amount of resources required by each task as well as other constraints such as the node where tasks should be deployed. Each task then aims to find a resource offer made by Mesos that matches its requirements. 

\subsubsection{Marathon}

Marathon~\cite{marathonOnline} is a meta-framework for Mesos that is designed to orchestrate long-running services. Because of this, it focuses on providing applications with fault-tolerance and high-availability; Marathon will ensure that launched applications will continue to run even in the presence of node failures. Aside from this, the framework also offers placement constraints, application health checks and monitoring, and application autoscaling. Pods as defined in Kubernetes (i.e., a set of applications that must be co-located) are supported in Marathon as of version 1.4. In this way, storage, networking, and other resources can be shared among multiple applications on a single node as defined by users. 

It is worthwhile mentioning that in terms of functionality, Marathon and Aurora are very similar products. However, there are a few differences. The main one is that Marathon handles only service-like jobs. Furthermore, setting up and using Marathon is considered to be simpler than doing so with Aurora~\cite{marathonVsAuroraOnline}; Marathon includes for example a user interface through which users can directly schedule tasks.  

\subsection{Apache Hadoop YARN}
YARN~\cite{vavilapalli2013apache} is a cluster manager designed to orchestrate Hadoop tasks, although it also supports other frameworks such as Giraph, Spark, and Storm. Each application framework running on top of YARN coordinates their own execution flows and optimizations as they see fit. In YARN, there is a per-cluster resource manager (RM) and an application master (AM) per framework. The application master requests resources from the resource manager and generates a physical plan from the resources it receives. The RM allocates containers to applications to run on specific nodes. In this context, a container is a bundle of resources bounded to run on a specific machine. There is one AM per job (a job is a set of tasks related to a framework) and it is responsible for managing its lifecycle. This includes increasing and decreasing resource consumption, managing the flow of executions, etc. The AM needs to harness the resources available on multiple nodes to complete a job. To obtain these, the AM requests resources from the RM and the request can include locality preferences and properties of the containers~\cite{li2016performance}. Finally, YARN supports two different containerizers, a custom built-in container manager based on Linux cgroups and Docker.

\subsection{Omega}
Omega~\cite{schwarzkopf2013omega} is Google's next generation cluster management system. As opposed to a monolithic or two-level scheduler as used by other approaches, Omega proposes the use of a parallel scheduler architecture built around shared state. In this way, Omega offers a platform that enables specialized and custom schedulers to be developed, providing users with a great deal of flexibility. The shared cluster state is maintained in a centralized, Paxos-based, transaction-oriented data store that is accessed by the different components of the architecture (such as schedulers). To handle conflicts derived from this, Omega uses an optimistic concurrency control approach. This means that occasionally a situation will arise in which two schedulers select the same set of resources for different tasks, hence, the scheduling of one of these tasks may have to be re-done. Despite this additional work, the overhead was found to be acceptable and the resulting benefits in eliminating blocking as would be done by a pessimistic concurrency approach was found to offer better scalability~\cite{schwarzkopf2013omega}.

\subsection{Apollo}
Apollo~\cite{boutin2014apollo} is a scheduling framework developed at Microsoft. It aims to balance scalability and scheduling quality by adopting a distributed and coordinated scheduling strategy. In this way, it avoids suboptimal decisions by independent distributed schedulers, and removes the scalability bottleneck of centralized ones. The scheduling of tasks is done so that the task completion time is minimized. This is a unique feature among all the surveyed systems. The runtime of tasks is estimated based on historical data statistics from similar tasks. Apollo uses opportunistic scheduling to drive high utilization while maintaining low job latencies. While regular tasks are scheduled to ensure low latency, opportunistic ones are scheduled to drive high utilization by filling the slack left by regular tasks. In Apollo, the physical execution plan of jobs is represented as DAGs, with the tasks representing a basic computation unit and the edges the data flow between tasks. Tasks of the same type are logically grouped together in stages, with the number of tasks per stage indicating the degree of parallelism of the DAG.

\subsection{Fuxi}
Fuxi~\cite{zhang2014fuxi} is a resource management and scheduling system that supports Alibaba's proprietary data platform. It is the resource management module on their Aspara system, which is responsible for managing the physical resources of Linux clusters within a data center and controlling the parallel execution of parallel applications. Users submit jobs to the FuxiMaster along with information such as the application type and the master package location. A FuxiAgent then launches the corresponding application master, which retrieves the application description and determines the resource demand for different stages of the job execution. The application master then sends resource requests to the FuxiMaster. When resources are granted, the application master sends concrete work plans to FuxiAgents. FuxiAgents use Linux cgroups to enforce resource constraints. When the application process finishes, the application master returns the resources back to FuxiMaster. Finally, incremental or locality-based scheduling enables Fuxi to make scheduling decisions in micro seconds. When resources are free, the decision of whom to allocate them to is only made between those applications in the machine's queue, as opposed to considering all the other existing machines and applications. 

\begin{table}[t]
\centering
\caption{Originating organizations of the surveyed systems}
\label{orgSurvey}
\begin{tabular}{@{}lllp{5cm}@{}}
\toprule
\textbf{System}     & \textbf{Originating Organization}  &\textbf{Open Source}  &{\textbf{Container Technology}}       \\ \midrule
Borg              & Google & \dash & Linux cgroups-based \\
Kubernetes           & Google &  \checkmark & Docker, rkt, CRI API implementations, OCI-compliant runtimes\\
Swarm              &  Docker & \checkmark & Docker\\
Mesos              &  UC Berkeley & \checkmark & Mesos containers, Docker\\
Aurora              &  Twitter & \checkmark & Mesos containers, Docker\\
Marathon              & Mesosphere & \checkmark & Mesos containers, Docker \\
YARN              &  Apache & \checkmark & Linux cgroups-based, Docker\\ 
Omega              & Google & \dash & N/S\\
Apollo              &  Microsoft & \dash & N/S \\
Fuxi              & Alibaba & \dash & Linux cgroups-based \\
\bottomrule
\end{tabular}
\end{table}

\begin{table}[t!]
\centering
\caption{System classification for the application model}
\label{appModelSurvey}
\begin{tabular}{@{}lllll@{}}
\toprule
\textbf{System}     & \textbf{Workload}  &\textbf{Job Composition} \\ \midrule
Borg              &  All     & Independent tasks\\
Kubernetes           &   All    & Co-located tasks \\
Swarm              & Long-running jobs   & Co-located tasks\\
Mesos              & All   & Single task\\
Aurora              &Long-running and cron jobs   & Independent tasks\\
Marathon          & Long-running jobs    & Co-located tasks\\
YARN             & Batch jobs  & Single task \\ 
Omega              & All     & Independent tasks\\
Apollo              & Batch jobs  & Task graph\\
Fuxi               & Batch jobs    & Task graph\\
\bottomrule
\end{tabular}
\end{table}

\begin{table}[t]
\centering
\caption{System classification for job scheduling}
\label{rschedSurvey}
\begin{tabular}{@{}lp{2.5cm}p{2.5cm}llp{2cm}@{}}
\toprule
\textbf{System}                      & \textbf{Architecture}  &\textbf{Node Selection} &\textbf{Preemption} &\textbf{Rescheduling} &\textbf{Placement Constraints} \\ \midrule
Borg              &  Centralized monolithic     & Randomized sample & \checkmark  &\checkmark &Value-based \\ 
Kubernetes              &    Decentralized monolithic   & All nodes &  \dash &\dash&Label and affinity-based \\
Swarm              &     Decentralized monolithic & All nodes &  \dash & \checkmark  &Label and affinity-based \\
Mesos              &     Two-level offer-based  & N/A &   \dash &\dash&N/A \\
Aurora              &      Two-level offer-based & All nodes &  \checkmark &\checkmark&Value and limit-based \\
Marathon              &     Two-level offer-based & All nodes &  \dash & \checkmark & Value and query-based \\
YARN              &      Two-level request based  & All nodes & \dash&\dash&Value and affinity-based \\
Omega              &     Decentralized modular  & All nodes &  \checkmark&\checkmark &N/S \\ 
Apollo              &    Decentralized monolithic   & Cluster partition and randomized sample &  \checkmark &\checkmark &None \\
Fuxi              &  Two-level Request-based   & All nodes & \checkmark& \checkmark &Value-based \\
\bottomrule
\end{tabular}
\end{table}

\begin{table}
\centering
\caption{System classification for cluster infrastructure}
\label{clusterInfSurvey}
\begin{tabular}{@{}p{2cm}p{4.5cm}p{4cm}@{}}
\toprule
\textbf{System}                   & \textbf{Cluster Elasticity}    & \textbf{Cluster Infrastructure}  \\ \midrule
Borg              &    Static & Non-virtualized    \\
Kubernetes              &  Elastic, manual and autoscaling  & Virtualized, non-virtualized   \\
Swarm              &      Elastic, manual scaling & Virtualized, non-virtualized \\
Mesos              &      Elastic, manual scaling & Virtualized, non-virtualized  \\
Aurora              &      Elastic,  manual scaling & Virtualized, non-virtualized  \\
Marathon              &     Elastic,  manual scaling & Virtualized, non-virtualized  \\
YARN              &       Elastic, manual scaling & Virtualized, non-virtualized   \\
Omega              &     Static &Non-virtualized   \\
Apollo              &       Static & Non-virtualized   \\
Fuxi              &      Static & Non-virtualized  \\
\bottomrule
\end{tabular}
\end{table}

\begin{table}[t!]
\centering
\caption{System classification for resource management}
\label{resMngmntSurvey}
\begin{tabular}{@{}p{1.7cm}p{2cm}p{2.5cm}p{2.2cm}p{1.7cm}p{1.7cm}@{}}
\toprule
\textbf{System}                &\textbf{Quota Management} &\textbf{Resource Reclamation} &\textbf{Resource Granularity} &\textbf{Oversubscrip- tion} &\textbf{Resource Estimation}\\ \midrule
Borg              & Limits, requests & Eviction, throttling & Fine-grained & \checkmark & \checkmark \\
Kubernetes              & Limits, requests & Eviction, throttling & Fine-grained & \checkmark & \dash \\
Swarm             & Requests & Eviction & Fine-grained & \dash &\dash  \\
Mesos              & Requests & Eviction, throttling & Fine-grained & \checkmark & \dash \\
Aurora              & Limits & Eviction, throttling & Fine-grained &\checkmark  & \dash \\
Marathon            & Requests & Eviction, throttling  & Fine-grained & \dash &\dash  \\
YARN              & Requests & Eviction & Coarse-grained & \dash & \dash \\
Omega             & N/S & N/S & Fine-grained &\checkmark & \dash  \\ 
Apollo              & Limits & Eviction, throttling & Fine-grained &\checkmark & \dash \\
Fuxi               & Requests & Eviction & Bundle &\dash  & \dash  \\
\bottomrule
\end{tabular}
\end{table}

\section{System Classification and Discussion}

This section contains the classification of the surveyed container orchestration systems based on the presented taxonomy. Firstly, as a reference to readers, Table \ref{orgSurvey} depicts the organization from which each of the surveyed systems originate from, as well as whether they are open source or proprietary and the container runtimes they support. Next, the classification of the studied systems is presented on Tables \ref{appModelSurvey}, \ref{rschedSurvey},  \ref{clusterInfSurvey},  \ref{resMngmntSurvey}, \ref{sysObjSurvey}, \ref{multitenancySecuritySurvey}, and \ref{multitenancyPerformanceSurvey}. Specifically, Table \ref{appModelSurvey} displays the application model summary while Table \ref{rschedSurvey} contains the classification from the scheduling perspective. Tables \ref{clusterInfSurvey}, \ref{resMngmntSurvey}, \ref{sysObjSurvey}, \ref{multitenancySecuritySurvey}, and \ref{multitenancyPerformanceSurvey} depict the classification of the studied systems from the cluster infrastructure and management perspective. Note that N/A is used when a property does not apply to the given system and N/S is used when details regarding the specific characteristic were not specified in the information sources that describe the system.

Overall, we found that the commercial systems include more advanced features than the open source ones. Some of these features such as preemption and rescheduling however, may be tailored for the specific needs of an organization and hence may be unsuitable for more general open source systems, unless the flexibility of deciding their goals is left up to users or developers. This is a complex endeavor. 

In terms of scheduling, two-level schedulers such as Mesos and YARN offer the most flexibility and extensibility by allowing each application framework to define their own scheduling policies based on their needs and specialized knowledge. However, framework schedulers in this type of systems are limited in the decisions they can make as they only have access to the cluster state information that is provided by the resource manager. To address this issue, shared-state schedulers such as Kubernetes and Omega support pluggable schedulers that can be used simultaneously to schedule different types of jobs with different characteristics. This approach may avoid the scalability and scheduler complexity issues of fully centralized schedulers such as Borg. However, the ability of the system to scale the number of schedulers to a large number while remaining efficient is an open research question. 

Oversubscribing resources seems a common practice in most systems with the aim of achieving high cluster utilization. However, how this is achieved in practice may have a considerable impact in the performance of applications and the orchestration system itself. For example, relying on users' estimates of an application resource requirements may not leave room for oversubscription or may make this process cumbersome in that resources are frequently exhausted (e.g., out of memory events) in servers and hence tasks have to be frequently evicted and rescheduled. Borg's approach to oversubscription seems to work well, with task resource usage being estimated based on fine grained resource consumption measurements over time, instead of relying solely on user resource requests. This approach raises another interesting feature that could greatly benefit container orchestration systems, resource consumption estimation.

All of the studied systems require users to specify the amount of resources, at least in terms of CPU and memory, that a task will consume. This is not only a challenging task for users but also a risk for orchestrators. Users may easily overestimate their resource requirements for example. This will lead to an inefficient use of resources with many remaining idle. Some work on automatically determining resource requirements would greatly facilitate the deployment of containerized applications for users, improve their quality of service, and allow cluster resources to be better utilized.

Regarding preemption, although a common denominator in proprietary systems, only Aurora supports this functionality in the open-source category. The use of preemption enables clusters to be used to run jobs with different purposes such as production or staging in a more robust manner. In this way for example, the resource manager can make room for production jobs by evicting staging ones if the cluster no longer has adequate resources. It may also work in the opposite way, with testing jobs opportunistically using resources but only if they are not in use or needed by production jobs.

Kubernetes was the only system with a cluster autoscaler, although it is not a built-in component but rather an add-on that is deployed as a pod in the cluster. The autoscaler's main goal is to place pods that failed to schedule due to insufficient resources in newly provisioned nodes. It is restricted however to provisioning only nodes that are similar in characteristics to those already in the cluster or node group. Hence, if a cluster is composed of \emph{small} VMs, the newly provisioned nodes will also be \emph{small}. Although this may achieve the purpose of placing a currently unschedulable node, the consequences in terms of cost and resource utilization for instance are not taken into consideration. 

Finally, the majority of the studied systems, except Swarm, were designed with multi-tenancy in mind. This is aligned with their goal of executing heterogeneous workloads on a set of shared resources. In particular, the frameworks are built with the aim of allowing multiple tenants from a single organization to share a single compute cluster. For the open source systems, information on different isolation features was mostly available. However, for the proprietary ones, the details on how isolation was achieved were elusive in the information sources. In particular, the classification for Omega, Apollo, and Fuxi, was partial due to the aforementioned reason. Despite this, it is clear from the use cases, problem formulation, scheduling approaches, and evaluation sections in the corresponding manuscripts that these frameworks do support intra-organizational multi-tenancy. In fact, for the specific case of Omega, it could be assumed that the multi-tenancy features of Borg apply to this system as well, but this could not be confirmed.

\begin{table}[t]
\centering
\caption{System classification for system objectives}
\label{sysObjSurvey}
\begin{tabular}{@{}p{1.7cm}p{1.8cm}p{2.1cm}p{2cm}p{2.2cm}p{2.1cm}@{}}
\toprule
\textbf{System}                      & \textbf{Scalability}  &\textbf{High Availability} &\textbf{High Utilization} &\textbf{High Throughput} &\textbf{Application QoS} \\ \midrule
Borg              &  \checkmark     & \checkmark & \checkmark  &\checkmark & \dash \\ 
Kubernetes              &  \checkmark     & \dash & \dash  & \dash & \dash \\ 
Swarm              &  \checkmark     & \checkmark & \checkmark & \checkmark & \dash \\ 
Mesos              &  \checkmark     & \checkmark & \dash  & \dash & \dash \\ 
Aurora             &  \checkmark     & \checkmark & \dash  & \dash & \dash \\ 
Marathon              &  \checkmark     & \checkmark & \dash  & \dash &\dash \\ 
YARN              &  \checkmark     & \checkmark  & \dash  & \checkmark & \dash \\ 
Omega              &  \checkmark  & \checkmark & \checkmark  & \checkmark & \dash \\ 
Apollo              &  \checkmark  & \dash & \checkmark  & \checkmark &\checkmark \\ 
Fuxi              &  \checkmark    & \checkmark & \checkmark  & \checkmark & \dash \\ 
\bottomrule
\end{tabular}
\end{table}

\begin{table}[h!]
\centering
\caption{Security isolation taxonomy}
\label{multitenancySecuritySurvey}
\begin{tabular}{@{}p{1.7cm}p{2cm}p{2cm}p{2.2cm}p{2cm}p{2cm}@{}}
    \toprule
                  &  \multicolumn{3}{c}{ \textbf{Access Control}}  &  \multicolumn{2}{c}{ \textbf{Container-centric}} \\
    \cmidrule(lr){2-4}
    \cmidrule(lr){5-6}
    \textbf{System}  &   \textbf{Restriced API Access}  &   \textbf{Authentication} &   \textbf{RBAC-based Authorization} &  \textbf{Restricted Execution} &  \textbf{Hypervisor Isolation} \\
    \midrule
Borg        & \checkmark & \checkmark  & N/S  & \dash  & \dash     \\
Kubernetes           & \checkmark & \checkmark & \checkmark & \checkmark & \checkmark   \\
Swarm             & \dash  & \checkmark &  \dash & \checkmark  & \dash     \\
Mesos            & \checkmark & \checkmark & \checkmark & \checkmark  & \dash   \\
Aurora             &  \checkmark &  \checkmark &  \checkmark & \checkmark & \dash   \\
Marathon          &  \checkmark &  \checkmark & \checkmark & \checkmark & \dash  \\
YARN           & \checkmark & \checkmark & \checkmark & \checkmark  & \dash    \\
Omega             &  N/S &  N/S & N/S & N/S & N/S   \\
Apollo              &  \checkmark  & \checkmark & \checkmark & N/S & N/S   \\
Fuxi              &  N/S & N/S & N/S & N/S & \dash  \\
    \bottomrule
  \end{tabular}
\end{table}

\begin{table}[t!]
\centering
\caption{Network and Performance isolation taxonomy}
\label{multitenancyPerformanceSurvey}
\begin{tabular}{@{}p{1.7cm}p{2.1cm}p{1.8cm}p{2.5cm}p{2.1cm}p{1.8cm}@{}}
\toprule
 & \multicolumn{3}{c}{ \textbf{Network Isolation}}  &  \multicolumn{2}{c}{ \textbf{Performance Isolation}} \\ 
   \cmidrule(lr){2-4}
    \cmidrule(lr){5-6}
    \textbf{System}  &   \textbf{IP-per-container}  &   \textbf{Port Mapping} &   \textbf{Third-party Plugins} &  \textbf{Compute Performance Isolation} &  \textbf{Per-tenant Resource Quota} \\
    \midrule
Borg              &  \dash     & \checkmark & \dash & \checkmark & \checkmark \\
Kubernetes       &   \checkmark    &  \dash & \checkmark & \checkmark & \checkmark \\
Swarm            &   \checkmark    &  \dash & \checkmark & \checkmark & \dash \\
Mesos             &    \checkmark   &  \checkmark & \checkmark & \checkmark & \checkmark \\
Aurora              &     \checkmark  & \checkmark & \checkmark & \checkmark & \checkmark\\
Marathon         &     \checkmark  & \checkmark & \checkmark & \checkmark & \checkmark \\
YARN            &    \dash   &  \dash & \dash & \checkmark & \checkmark \\
Omega             &    N/S   &  N/S & N/S & \checkmark & N/S \\
Apollo            &     N/S  & N/S  & N/S & \checkmark & \checkmark\\
Fuxi                 &  N/S     & N/S & N/S & \checkmark & \checkmark\\
\bottomrule
\end{tabular}
\end{table}

\section{Future Directions}
\label{futureDirections}
Although most systems (mostly the proprietary ones) are mature and include advanced features, the optimization space can still be further explored. This is especially true in the era of cloud computing, as most existing frameworks ignore many of the inherent features of cloud computing in favor of assuming a static cluster of resources. As a result, elasticity, resource costs, and pricing and service heterogeneities are ignored. For organizations deploying their workloads through container orchestrators in cloud, this translates into higher and unnecessary costs, potentially reduced application performance, and a considerable amount of man hours in tuning their virtual cluster to meet their needs. 

In light of this, a possible optimization to current systems is related to rescheduling. In particular, rescheduling for either defragmentation or autoscaling when the workload includes long-running tasks. Regardless of how good the initial placement of these tasks is, it will degrade over time as the workload changes. This will lead to an inefficient use of resources in which the load is thinly spread across nodes or the amount of resources in different nodes are not sufficient to run other applications. Rescheduling applications that tolerate a component being shut down and restarted will enable the orchestration system to consolidate and rearrange tasks so that more applications can be deployed on the same number of nodes or some nodes can be shutdown to reduce cost or save energy. Similarly, if more nodes are added to the cluster, being able to reschedule some of the existing applications on the new nodes may be beneficial in the long term.

Another future direction is for cloud-aware placement algorithms to consider the heterogeneities of the underlying resources, including different pricing models, locations, and resource types and sizes. This would enable for instance to dynamically provision resources of different pricing models to the virtual cluster in order to satisfy growing needs of the applications with minimum cost. For example, a customer-facing application should be placed on reserved instances that are leased for lower costs and longer periods of time while offering high availability. Batch jobs on the other hand could be placed on unreliable rebated resources, whose sudden termination will not disrupt the end user experience. The use of on-demand instances can be explored for applications with requirements in between where the availability is needed but they are not long-running services. To realize these goals, it is required to filter unqualified resources and propose new resource affinity models to rank the resources when provisioning for different applications. These policies can be implemented as extensions of the existing filtering and affinity ranking mechanisms of the current platforms for example.

Even though intra-organizational multi-tenancy is the most common use case for containerized clusters, existing frameworks are continuously striving to facilitate stronger isolation mechanisms to suit stricter multi-tenancy requirements that satisfy the requirements of inter-organizational multi-tenancy. For example, enabling tenants to have their own control plane objects is yet to be a feature of existing systems. Another important consideration is related to fault isolation; although frameworks offer several failure management and recovery mechanisms to applications, work is still required to ensure that a failure from one tenant does not cascade to other tenants. For example, the failure of a node's operating system due to a faulty or malicious container will cause other containers sharing that node to fail. The consequences of this may differ for different application types. For instance, those with replicated containers may not be highly impacted. However, non-replicated critical applications may suffer grave consequences from such an event. Although this may be an inevitable side-effect of OS-level virtualization, the risks can potentially be mitigated for example by scheduling critical applications or those that are not replicated in single-tenant nodes. Another important aspect to consider in terms of security are denial of service attacks, preventing or mitigating the effects of these within a cluster or a network of containers is still in need of research. 

Finally, application QoS management is limited in existing systems. It is not unusual for applications to have specific QoS requirements. For instance, long-running services commonly have to serve a minimum number of requests per time unit or have stringent latency requirements. Batch jobs on the other hand can have a deadline as a time constraint for their execution or may need to be completed as fast as possible. For the first scenario, many systems offer a basic autoscaling mechanism. It monitors the CPU utilization of a service, and if a predefined threshold is exceeded, another instance of the service is launched. This however, is a baseline approach to autoscaling and integrating more sophisticated approaches to container-based management systems is required. For batch jobs, orchestrating them and assigning them to resources so that their QoS are met is another open research area. Although Apollo addresses this to a certain degree, this feature is not present in any open source system and support for heterogeneous QoS constraints is still unexplored. 


\section{Summary and Conclusions}
\label{summary}
In this work, we studied orchestration systems that are designed to manage the deployment of containerized applications in large-scale clusters. The growing popularity of container technologies has been a driving force contributing to the evolution and increased adoption of these systems in recent years. Frameworks such as Kubernetes are being widely used by organizations to deploy their large-scale workloads that include diverse applications such as web services and big data analytics. They are designed to manage the deployment of applications in clusters and are capable of running hundreds of thousands of jobs across thousands of machines. To achieve this, they are designed to address important challenges such as scalability, fault-tolerance and availability, efficient resource utilization, and request throughput maximization among others. 

To better understand containerized management systems, this work first introduced a reference architecture identifying key components and their responsibilities. These were segregated in four hierarchical layers, namely an application layer composed of jobs submitted by users, a cluster manager master in charge of orchestrating these jobs and managing the cluster resources, a cluster of worker compute nodes, and the infrastructure where the nodes are deployed. Based on these layers, the roles of the components within them, and existing frameworks, a taxonomy identifying various characteristics of container-based cluster management systems from the perspective of their application, scheduling, and cluster management models was proposed. Classifications based on the types of workload supported, the features supporting multi-tenancy, the scheduler architecture and scheduling policies, the elasticity of the cluster, the management of nodes and their resources, and the system's objectives were proposed and discussed in detail. A survey of the state state-of-the-art systems was also presented, and the proposed taxonomy applied to them.

Furthermore, future directions derived from gaps identified in the literature were presented with the aim of guiding emerging research. In particular, we identified the need for further work exploring elastic, cloud-based clusters. This would encompass addressing issues such as cost-aware scheduling that consider the heterogeneity of cloud resources such as different pricing models, geographical locations, and VM costs and specifications. Rescheduling to address not only defragmentation but to support the efficient use of a dynamic cluster achieved through autoscaling is another topic that requires further attention. Managing the QoS requirements of applications is another area that should be further explored. For example, guaranteeing the execution time of batch jobs to be within a specified deadline is yet to be a feature of any of the surveyed open source systems. Finally, extending the functionality of existing frameworks to provided increased isolation features supporting various multi-tenancy use cases is necessary. Although the vast majority of studied frameworks successfully support intra-organizational multi-tenancy use cases, further research is required to enable them to provide the foundation for models such as container as a service for example.

To conclude, management systems orchestrating containerized jobs in clusters are growing in popularity. Their adoption to provide as-a-service models resembling those offered by cloud computing as well as their use within organizations will continue to increase. This due to their ease of use and flexibility, their ability to efficiently use resources, their performance offerings, and advances in container technologies, among others. It is important then, for these systems to continue to be developed and optimized to offer users with varying requirements a robust solution to their needs and aid in shaping the future of distributed computing and applications. \\

\bibliography{references}

\begin{thebibliography}{46}
\providecommand{\natexlab}[1]{#1}
\providecommand{\url}[1]{\texttt{#1}}
\providecommand{\urlprefix}{}

\bibitem[{{Docker}(2018)}]{dockerOnline}
{Docker}, {Docker}; 2018.
\newblock \url{https://www.docker.com/}. Accessed June 2018.

\bibitem[{{Linux Containers}(2018)}]{lxcOnline}
{Linux Containers}, {LXC}; 2018.
\newblock \url{https://linuxcontainers.org/lxc/}. Accessed June 2018.

\bibitem[{{OpenVZ}(2018)}]{openvzOnline}
{OpenVZ}, {OpenVZ}; 2018.
\newblock \url{ http://openvz.org/}. Accessed June 2018.

\bibitem[{{Linux VServer}(2018)}]{vserverOnline}
{Linux VServer}, {Linux VServer}; 2018.
\newblock \url{ http://linux-vserver.org/}. Accessed June 2018.

\bibitem[{{rkt}(2018)}]{rktOnline}
{rkt}, {rkt}; 2018.
\newblock \url{https://coreos.com/rkt/}. Accessed June 2018.

\bibitem[{Felter et~al.(2015)Felter, Wes and Ferreira, Alexandre and Rajamony,
  Ram and Rubio, Juan}]{felter2015updated}
Felter W, Ferreira A, Rajamony R, Rubio J.
\newblock An updated performance comparison of virtual machines and linux
  containers.
\newblock In: Proceedings of the IEEE International Symposium On Performance
  Analysis of Systems and Software (ISPASS) IEEE; 2015. p. 171--172.

\bibitem[{Morabito et~al.(2015)Morabito, Roberto and Kj{\"a}llman, Jimmy and
  Komu, Miika}]{morabito2015hypervisors}
Morabito R, Kj{\"a}llman J, Komu M.
\newblock Hypervisors vs. lightweight virtualization: a performance comparison.
\newblock In: Proceedings of the IEEE International Conference on Cloud
  Engineering (IC2E) IEEE; 2015. p. 386--393.

\bibitem[{Ruiz et~al.(2015)Ruiz, Cristian and Jeanvoine, Emmanuel and Nussbaum,
  Lucas}]{ruiz2015performance}
Ruiz C, Jeanvoine E, Nussbaum L.
\newblock Performance evaluation of containers for {HPC}.
\newblock In: European Conference on Parallel Processing Springer; 2015. p.
  813--824.

\bibitem[{Piraghaj et~al.(2017)Piraghaj, Sareh Fotuhi and Dastjerdi, Amir Vahid
  and Calheiros, Rodrigo N and Buyya, Rajkumar}]{piraghaj2017containercloudsim}
Piraghaj SF, Dastjerdi AV, Calheiros RN, Buyya R.
\newblock ContainerCloudSim: An environment for modeling and simulation of
  containers in cloud data centers.
\newblock Software: Practice and Experience 2017;47(4):505--521.

\bibitem[{Jennings and Stadler(2015)Jennings, Brendan and Stadler,
  Rolf}]{jennings2015resource}
Jennings B, Stadler R.
\newblock Resource management in clouds: Survey and research challenges.
\newblock Journal of Network and Systems Management 2015;23(3):567--619.

\bibitem[{Manvi and Shyam(2014)Manvi, Sunilkumar S and Shyam, Gopal
  Krishna}]{manvi2014resource}
Manvi SS, Shyam GK.
\newblock Resource management for Infrastructure as a Service (IaaS) in cloud
  computing: A survey.
\newblock Journal of Network and Computer Applications 2014;41:424--440.

\bibitem[{Mann(2015)Mann, Zolt{\'a}n {\'A}d{\'a}m}]{mann2015allocation}
Mann Z{\'A}.
\newblock Allocation of virtual machines in cloud data centers - a survey of
  problem models and optimization algorithms.
\newblock ACMComputing Surveys (CSUR) 2015;48(1):11.

\bibitem[{Verma et~al.(2015)Verma, Abhishek and Pedrosa, Luis and Korupolu,
  Madhukar and Oppenheimer, David and Tune, Eric and Wilkes,
  John}]{verma2015large}
Verma A, Pedrosa L, Korupolu M, Oppenheimer D, Tune E, Wilkes J.
\newblock Large-scale cluster management at Google with Borg.
\newblock In: Proceedings of the 10th European Conference on Computer Systems
  ACM; 2015. p.~18.

\bibitem[{{Kubernetes}(2018)}]{kubernetesOnline}
{Kubernetes}, Kubernetes; 2018.
\newblock \url{https://kubernetes.io/}. Accessed June 2018.

\bibitem[{Boutin et~al.(2014)Boutin, Eric and Ekanayake, Jaliya and Lin, Wei
  and Shi, Bing and Zhou, Jingren and Qian, Zhengping and Wu, Ming and Zhou,
  Lidong}]{boutin2014apollo}
Boutin E, Ekanayake J, Lin W, Shi B, Zhou J, Qian Z, et~al.
\newblock Apollo: Scalable and Coordinated Scheduling for Cloud-Scale
  Computing.
\newblock In: Proceedings of the USENIX Symposium on Operating Systems Design
  and Implementation (OSDI); 2014. p. 285--300.

\bibitem[{Casavant and Kuhl(1988)Casavant, Thomas L. and Kuhl, Jon
  G.}]{casavant1988taxonomy}
Casavant TL, Kuhl JG.
\newblock A taxonomy of scheduling in general-purpose distributed computing
  systems.
\newblock IEEE Transactions on Software Engineering 1988;14(2):141--154.

\bibitem[{Toptal and Sabuncuoglu(2010)Toptal, Ay{\c{s}}eg{\"u}l and
  Sabuncuoglu, Ihsan}]{toptal2010distributed}
Toptal A, Sabuncuoglu I.
\newblock Distributed scheduling: a review of concepts and applications.
\newblock International Journal of Production Research 2010;48(18):5235--5262.

\bibitem[{Krauter et~al.(2002)Krauter, Klaus and Buyya, Rajkumar and
  Maheswaran, Muthucumaru}]{krauter2002taxonomy}
Krauter K, Buyya R, Maheswaran M.
\newblock A taxonomy and survey of grid resource management systems for
  distributed computing.
\newblock Software: Practice and Experience 2002;32(2):135--164.

\bibitem[{Schwarzkopf et~al.(2013)Schwarzkopf, Malte and Konwinski, Andy and
  Abd-El-Malek, Michael and Wilkes, John}]{schwarzkopf2013omega}
Schwarzkopf M, Konwinski A, Abd-El-Malek M, Wilkes J.
\newblock Omega: flexible, scalable schedulers for large compute clusters.
\newblock In: Proceedings of the 8th ACM European Conference on Computer
  Systems ACM; 2013. p. 351--364.

\bibitem[{Hindman et~al.(2011)Hindman, Benjamin and Konwinski, Andy and
  Zaharia, Matei and Ghodsi, Ali and Joseph, Anthony D and Katz, Randy H and
  Shenker, Scott and Stoica, Ion}]{hindman2011mesos}
Hindman B, Konwinski A, Zaharia M, Ghodsi A, Joseph AD, Katz RH, et~al.
\newblock Mesos: A Platform for Fine-Grained Resource Sharing in the Data
  Center.
\newblock In: Proceedings of the USENIX Symposium on Networked Systems Design
  and Implementation (NSDI); 2011. p. 22--22.

\bibitem[{Zhang et~al.(2014)Zhang, Zhuo and Li, Chao and Tao, Yangyu and Yang,
  Renyu and Tang, Hong and Xu, Jie}]{zhang2014fuxi}
Zhang Z, Li C, Tao Y, Yang R, Tang H, Xu J.
\newblock Fuxi: a fault-tolerant resource management and job scheduling system
  at internet scale.
\newblock Proceedings of the VLDB Endowment 2014;7(13):1393--1404.

\bibitem[{{Marathon}(2018)}]{marathonOnline}
{Marathon}, Marathon; 2018.
\newblock \url{https://mesosphere.github.io/marathon/}. Accessed June 2018.

\bibitem[{{Apache Aurora}(2018)}]{apacheAuroraOnline}
{Apache Aurora}, Apache Aurora; 2018.
\newblock \url{http://aurora.apache.org/}. Accessed June 2018.

\bibitem[{Vavilapalli et~al.(2013)Vavilapalli, Vinod Kumar and Murthy, Arun C
  and Douglas, Chris and Agarwal, Sharad and Konar, Mahadev and Evans, Robert
  and Graves, Thomas and Lowe, Jason and Shah, Hitesh and Seth, Siddharth and
  others}]{vavilapalli2013apache}
Vavilapalli VK, Murthy AC, Douglas C, Agarwal S, Konar M, Evans R, et~al.
\newblock Apache hadoop yarn: Yet another resource negotiator.
\newblock In: Proceedings of the 4th annual Symposium on Cloud Computing ACM;
  2013. p.~5.

\bibitem[{{Kata Containers}(2018)}]{kataOnline}
{Kata Containers}, {Kata Containers}; 2018.
\newblock \url{https://katacontainers.io/}. Accessed August 2018.

\bibitem[{{runV}(2018)}]{runvOnline}
{runV}, {runV - bring Isolation to Docker}; 2018.
\newblock \url{https://blog.hyper.sh/runv-bring-isolation-to-docker.html}.
  Accessed August 2018.

\bibitem[{{vSphere Integrated Containers}(2018)}]{vsphereOnline}
{vSphere Integrated Containers}, {vSphere Integrated Containers}; 2018.
\newblock \url{https://vmware.github.io/vic-product/}. Accessed August 2018.

\bibitem[{{Project Calico}(2018)}]{calicoOnline}
{Project Calico}, {Calico}; 2018.
\newblock \url{https://www.projectcalico.org/}. Accessed August 2018.

\bibitem[{{CoreOS}(2018)}]{flannelOnline}
{CoreOS}, {Flannel}; 2018.
\newblock \url{https://coreos.com/flannel/docs/latest/}. Accessed August 2018.

\bibitem[{{Nuage}(2018)}]{nuageOnline}
{Nuage}, {Nuage}; 2018.
\newblock \url{http://www.nuagenetworks.net/}. Accessed August 2018.

\bibitem[{{Linux}(2018)}]{cgroupsOnline}
{Linux}, cgroups - Linux Control Groups; 2018.
\newblock \url{http://man7.org/linux/man-pages/man7/cgroups.7.html/}. Accessed
  June 2018.

\bibitem[{{Google}(2018)}]{lmctfyOnline}
{Google}, lmctfy; 2018.
\newblock \url{https://opensource.google.com/projects/lmctfy/}. Accessed August
  2018.

\bibitem[{Burrows(2006)Burrows, Mike}]{burrows2006chubby}
Burrows M.
\newblock The Chubby lock service for loosely-coupled distributed systems.
\newblock In: Proceedings of the 7th symposium on Operating systems design and
  implementation USENIX Association; 2006. p. 335--350.

\bibitem[{Melnik et~al.(2010)Melnik, Sergey and Gubarev, Andrey and Long, Jing
  Jing and Romer, Geoffrey and Shivakumar, Shiva and Tolton, Matt and
  Vassilakis, Theo}]{melnik2010dremel}
Melnik S, Gubarev A, Long JJ, Romer G, Shivakumar S, Tolton M, et~al.
\newblock Dremel: interactive analysis of web-scale datasets.
\newblock Proceedings of the VLDB Endowment 2010;3(1-2):330--339.

\bibitem[{Lamport(1998)Lamport, Leslie}]{lamport1998part}
Lamport L.
\newblock The part-time parliament.
\newblock ACM Transactions on Computer Systems (TOCS) 1998;16(2):133--169.

\bibitem[{{Containerd}(2018)}]{containerdOnline}
{Containerd}, Containerd; 2018.
\newblock \url{https://containerd.io/}.Accessed August 2018.

\bibitem[{{Kubernetes}(2018)}]{fraktiOnline}
{Kubernetes}, Frakti; 2018.
\newblock \url{https://github.com/kubernetes/frakti/}. Accessed August 2018.

\bibitem[{{CRI-O}(2018)}]{crioOnline}
{CRI-O}, {CRI-O}; 2018.
\newblock \url{http://cri-o.io/}. Accessed August 2018.

\bibitem[{{Opencontainers runc}(2018)}]{runcOnline}
{Opencontainers runc}, {runc}; 2018.
\newblock \url{https://github.com/opencontainers/runc/}. Accessed August 2018.

\bibitem[{{CoreOS}(2018)}]{etcdOnline}
{CoreOS}, etcd; 2018.
\newblock \url{https://coreos.com/etcd/}. Accessed June 2018.

\bibitem[{{Kubernetes}(2018)}]{kubernetesEvalOnline}
{Kubernetes}, Building Large Clusters; 2018.
\newblock \url{https://kubernetes.io/docs/admin/cluster-large/}. Accessed June
  2018.

\bibitem[{{Docker}(2018)}]{dockerSwarmOnline}
{Docker}, Docker Swarm; 2018.
\newblock \url{https://docs.docker.com/engine/swarm/}. Accessed June 2018.

\bibitem[{{ Docker Inc.}(2018)}]{swarmEval}
{ Docker Inc }, Scale Testing Docker Swarm to 30,000 Containers; 2018.
\newblock
  \url{https://blog.docker.com/2015/11/scale-testing-docker-swarm-30000-containers}.
  Accessed June 2018.

\bibitem[{Hunt et~al.(2010)Hunt, Patrick and Konar, Mahadev and Junqueira,
  Flavio Paiva and Reed, Benjamin}]{hunt2010zookeeper}
Hunt P, Konar M, Junqueira FP, Reed B.
\newblock ZooKeeper: Wait-free Coordination for Internet-scale Systems.
\newblock In: Proceedings of the USENIX annual technical conference, vol.~8
  Boston, MA, USA; 2010. .

\bibitem[{{Mohit Gupta}(2018)}]{marathonVsAuroraOnline}
{Mohit Gupta}, Marathon vs Aurora: A High Level Comparison; 2018.
\newblock
  \url{http://code.hootsuite.com/marathon-vs-aurora-a-high-level-comparison/}.
  Accessed June 2018.

\bibitem[{Li et~al.(2016)Li, Bo and Song, Meina and Ou, Zhonghong and Haihong,
  E}]{li2016performance}
Li B, Song M, Ou Z, Haihong E.
\newblock Performance Comparison and Analysis of Yarn's Schedulers with Stress
  Cases.
\newblock In: Proceedings of the 7th International Conference on Cloud
  Computing and Big Data (CCBD) IEEE; 2016. p. 93--98.

\end{thebibliography}

\end{document}